\documentclass[12pt]{article}
\usepackage{amsmath}
\usepackage{latexsym}
\usepackage[dvips]{graphicx}
\usepackage{amssymb}
\usepackage{epsfig}

\newif{\ifcomentarios}
\comentariosfalse

\newtheorem{theorem}{Theorem}

\newtheorem{proposition}[theorem]{Proposition}

\renewcommand{\Bbb}{\mathbb}

\begin{document}

\author{{\bf Domingos H. U. Marchetti}\thanks{%
Partially supported by FAPESP and CNPq. {\bf E-mail:} marchett@ime.usp.br} \
\& \ {\bf Claudio F. de Souza Rodrigues}\thanks{%
Supported by FAPESP under grant 98/06804-2R. {\bf E-mail:} claudiof@if.usp.br%
} \\
Instituto de F\'{\i}sica \\
Universidade de S\~ao Paulo\\
Caixa Postal 66318\\
05315-970 S\~ao Paulo, Brasil\\
}
\title{COMBINATORIAL SOLUTION OF ONE--DIMENSIONAL QUANTUM SYSTEMS}
\date{}
\maketitle

\begin{abstract}
We give a self--contained exposition of the combinatorial solution of
quantum mechanical systems of coupled spins on a one--dimensional lattice.
Using Trotter formula, we write the partition function as a generating
function of a spanning subgraph of a two--dimensional lattice and solve the
combinatorial problem by the method of Pfaffians provided the weights
satisfy the so--called free fermion condition. The free energy $f=f(\beta ,%
{\bf J},{\bf h})$ and the ground state energy $e_{0}=e_{0}({\bf J},{\bf h})$
as a function of the inverse temperature $\beta $, couplings ${\bf J}$ and
magnetic fields ${\bf h}$, for the $XY$ model in a transverse field with
period $p=1$ and $2$, is then obtained.
\end{abstract}



\section{Introduction}

\setcounter{equation}{0} \setcounter{theorem}{0}

A quantum mechanical system of coupled spins on a one-dimensional lattice $%
\Lambda \in {\Bbb Z}$ with $|\Lambda |=N$ sites is described by Hamiltonians
of the form
\begin{equation}
H_{N}(\sigma )=-\sum_{i\in \Lambda }h_{i,i+1}  \label{H}
\end{equation}
where 
\begin{equation}
h_{i,j}=J_{i,j}^{x}\;\sigma _{i}^{x}\,\sigma _{j}^{x}+J_{i,j}^{y}\;\sigma
_{i}^{y}\,\sigma _{j}^{y}+J_{i,j}^{z}\;\sigma _{i}^{z}\,\sigma _{j}^{z}+%
\frac{h_{i}}{2}(\sigma _{i}^{z}+\sigma _{j}^{z})\;.  \label{h}
\end{equation}
For each $i\in \Lambda $, $\sigma _{i}^{\alpha },\alpha =x,y,z$, are Pauli
matrices and periodic boundary condition is imposed, $\sigma _{i+N}^{\alpha
}=\sigma _{i}^{\alpha }$. The family of couplings ${\bf J}%
=\{(J_{i,i+1}^{x},J_{i,i+1}^{y},J_{i,i+1}^{z})\}_{i\in \Lambda }$ and
magnetic fields ${\bf h}=\{h_{i}\}_{i\in \Lambda }$ we shall considered here
are assumed to be periodic of period $p$ and, for simplicity, only $p=1$, $%
J_{i,i+1}^{\alpha }=J^{\alpha }$, $h_{i}=h$ and $p=2$ arrays, $%
J_{i,i+1}^{\alpha }=J_{1}^{\alpha }$, $h_{i}=h_{1}$ if $i$ is odd and $%
J_{i,i+1}^{\alpha }=J_{2}^{\alpha }$, $h_{i}=h_{2}$ if $i$ is even, will be
treated. The $z$--component couplings $J_{i,i+1}^{z}$ will also be set iqual
to zero at certain moment.

One-dimensional quantum systems have been extensively studied, as compared
to the higher dimensional counterparts, because some of their physical
quantities can be explicitly calculated. In the present work we consider the
free energy function, $f=f(\beta ,{\bf J},{\bf h})$, and the ground state
energy, $e_{0}=e_{0}({\bf J},{\bf h})=\displaystyle\lim_{\beta \rightarrow
\infty }f$, where 
\begin{equation}
f=\lim_{N\rightarrow \infty }\frac{-1}{\beta N}\ln Z_{N}\,.  \label{fe}
\end{equation}
The partition function 
\begin{equation}
Z_{N}={\rm Tr}\;e^{-\beta \,H_{N}}\equiv {\rm Tr}\;\rho _{N}\;,  \label{Z}
\end{equation}
is given by the trace over a density matrix $\rho _{N}$, defined in the $%
2^{N}$--dimensional vector space $\bigotimes_{i\in \Lambda }{\Bbb C}^{2}$,
and $\beta $ is the inverse temperature.

Among the available solution methods, we distinguish two types depending on
whether one approaches the partition function $Z_{N}$ directly or by means
of the eigenvalue problem of $H_{N}$.

The eigenvalues are more easily accessible when, by a canonical
transformation, $H_{N}$ can be written as a free fermion system. If ${\bf J}$
and ${\bf h}$ have period one with $J^{z}=0$, then (\ref{H}) is unitarily
equivalent to \cite{K} 
\begin{equation}
{\cal H}_{N}=-t\sum_{i\in \Lambda }\left\{ c_{i}^{\dagger
}\,c_{i+1}-c_{i}\,c_{i+1}^{\dagger }+\Gamma \left( c_{i}^{\dagger
}\,c_{i+1}^{\dagger }-c_{i}\,c_{i+1}\right) +H\left( 1-2\,c_{i}^{\dagger
}\,c_{i}\right) \right\} \,,  \label{H1}
\end{equation}
where $c_{i}^{\dagger }$ and $c_{i}$ are the creation and annihilation
operators for electrons at the site $i\in \Lambda $, $t=J^{x}+J^{y}$, $%
\Gamma =(J^{x}-J^{y})/t$, $H=h/t$ and the boundary condition terms has been
omitted. Lieb {\em et al} \cite{LSM} and Katsura \cite{Ka} have calculated
the spectrum of the $XY$--model ((\ref{H1}) with $H=0$) and $XY$--model in a
transverse field (\ref{H1}), respectively, obtaining the free energy
function and the ground state. Subsequently, the $XY$--model with
alternating interactions and magnetic moments has been solved by Perk {\em %
et al} \cite{PCZ}.

The partition function is said to be approached directly if it is mapped
into a problem of counting spanning subgraphs in a lattice. Direct approach
does not require $H_{N}$ to be diagonalizable but in order to solve the
underlying problem certain conditions have to be met. The basic ingredient
for this approach is the Trotter formula \cite{T}, 
\begin{equation}
e^{\left( K^{1}+K^{2}\right) }=\lim_{m\rightarrow \infty }\left(
e^{K^{1}/m}\,e^{K^{2}/m}\right) ^{m}\;,  \label{Trotter}
\end{equation}
valid for any finite-dimensional matrices $K^{1},K^{2}$ (see e.g. \cite{S}
for extensions and applications). The partition function (\ref{Z}) can thus
be mapped into a classical two dimensional lattice spin system as explained
in detail in the following sections.

There is another relationship between chains of quantum spins and the
transfer matrix of two--dimensional classical systems which has been
discussed in the excellent review by Kasteleyn \cite{K1}. 

The partition function has been directly
approached by Suzuki and Inoue \cite{SuI} who, using Trotter formula, have
mapped the {\em period--one} $XY$--model in a transverse field into a {\em %
period--two} $8$--vertex model with two sets of weights, $%
\{w_{i}\}_{i=1}^{8} $ and $\{w_{i}^{\prime }\}_{i=1}^{8}$, satisfying the
free--fermion condition 
\begin{equation}
w_{1}w_{2}+w_{3}w_{4}=w_{5}w_{6}+w_{7}w_{8}\,.  \label{ff}
\end{equation}
Equation (\ref{ff}) is the very fact that allowed Fan and Wu \cite{FW} to
solve the $8$--vertex model by the Pfaffian method (see \cite{HLW} for the
period $2$ model).

The purpose of this paper is to give a step--by--step account of the
combinatorial method of solving chains of interacting quantum spins. In
particular, Katsura's results are shown to be recovered by mapping the $XY$%
--model into a {\em period--one} $8$--vertex model with weight function
satisfying (\ref{ff}). For this we use the one--to--one mapping established
by Barma and Shastry \cite{BS} in the context of $XYZ$--model. Due to the
careful treatment of our exposition the method is extended to any
period $p$. To our knowledge, we calculate for the first time the free
energy of a  {\em period--two} $XY$--model in a transverse magnetic
field by the method of Pfaffian.

This paper is part of a program aiming at alternative solutions to the
ground state energy of Heisenberg (\ref{H}) models. The method based in the
spectral analysis of $H_{N}$ is amazingly far more developed than the direct
approach. The ground state energy of Hubbard \cite{LW} and Heisenberg models 
\cite{YY} can be obtained by an Ansatz to the eigenfunctions of $H_{N}$
called Bethe--Ansatz. The need for alternative solutions is more demanding
since the ``string hypothesis'' (completeness of the Bethe--Ansatz
eigenstates) remains unresolved.

Last but not least, the interest on the Pfaffian method has been renewed in
view of a recent result by Pinson and Spencer \cite{PS} on the universality
of critical $d=2$ Ising model.

The layout of this paper is as follows. In Section \ref{TF} the Hamiltonian
of the quantum spin system (\ref{H}) is mapped into a two--dimensional
classical spin system. This in turn is mapped, in Section \ref{vertex}, into
a spanning subgraph problem which is solved in Section \ref{pfaffian} by the
Pfaffian method. The evaluation of the determinant is presented in Section 
\ref{det} and free energy functions for periodic systems with periods $p=1$
and $2$ are determined in Sections \ref{FE} and \ref{XY-2}, respectively.


\section{Trotter Formula}

\label{TF} \setcounter{equation}{0} \setcounter{theorem}{0}

In this section the Trotter formula will be used to write the partition
function (\ref{Z}) as the partition function of a two-dimensional lattice $%
{\Bbb Z}^{N}\times {\Bbb Z}^{m}$ of a classical spin system in the limit $%
m\rightarrow \infty $.

Trotter's product formula (\ref{Trotter}) is a mathematical tool which has
elucidated the ``path integration'' of a quantum mechanical particle in a
potential (see \cite{S} and references therein for an accountable survey).
The application of this formula for coupled quantum spin systems was
originally carried out by Suzuki \cite{Su} but we shall here follow a
slightly modification due to Barma--Shastry \cite{BS}. Before working it
out, let us define the Hamiltonian system precisely.

To each site $i\in \Lambda $ it is associated a Hilbert space ${\cal H}_{i}=
{\Bbb C}^{2}$. The vectors $e_{+}=\left( {{1}\atop{0}}\right) $ and $
e_{-}=\left( {{0}\atop{1}}\right) $ form a base of ${\cal H}_{i}$ and
represent the two possible states of a spin $s_{i}$: $s_{i}=+$ if the spin
at $i$ is up and $s_{i}=-$ if it is down.

The vector space ${\cal H}={\cal H}_{\Lambda }$ where the Hamiltonian
operator (\ref{H}) acts is a tensor product space, 
\begin{equation}
{\cal H}=\bigotimes_{i\in \Lambda }{\cal H}_{i}\;.  \label{PH}
\end{equation}
Under the above interpretation, each of the $2^{N}$ vectors of the form 
\[
\left( {{1}\atop{0}}\right) \otimes \left( {{0}\atop{1}}\right) \otimes
\cdots \otimes \left( {{1}\atop{0}}\right) 
\]
represents a possible spin configuration ${\bf s}=\{s_{i}\}_{i\in \Lambda }$
of the whole system. Note that the collection ${\cal B}={\cal B}_{\Lambda
}=\{e_{{\bf s}}\}$ of vectors $e_{{\bf s}}:=e_{s_{1}}\otimes
e_{s_{2}}\otimes \cdots \otimes e_{s_{N}}$ with ${\bf s}\in \{+,-\}^{N}$
forms an orthonormal base of ${\cal H}$. So, if $E_{{\bf s}}=e_{{\bf s}
}\,(e_{{\bf s}})^{T}$ denotes the (orthogonal) projector of ${\cal H}$ into
the $e_{{\bf s}}$ direction, we have 
\begin{equation}
\sum_{{\bf s}\in \{+,-\}^{N}}E_{{\bf s}}={\it I\!\!\!\!I}\,,  \label{UM}
\end{equation}
where ${\it I\!\!\!\!I}=I\otimes I\otimes \cdots \otimes I$ is the identity
matrix: ${\it I\!\!\!\!I}v=v$, for all $v\in {\cal H}$.

A base of operators in ${\Bbb C}^{2}$ can be chosen to be the Pauli matrices 
\begin{equation}
\sigma ^{x}=\left( 
\begin{array}{cc}
0 & 1 \\ 
1 & 0
\end{array}
\right) \;,\;\;\;\;\;\;\;\;\;\sigma ^{y}=\left( 
\begin{array}{cc}
0 & -i \\ 
i & 0
\end{array}
\right) \;,\;\;\;\;\;\;\;\;\;\;\sigma ^{z}=\left( 
\begin{array}{cc}
1 & 0 \\ 
0 & -1
\end{array}
\right) \,,  \label{Pauli}
\end{equation}
and the identity $I=\left( 
\begin{array}{cc}
1 & 0 \\ 
0 & 1
\end{array}
\right) $. For later reference, we shall also consider the ``raising'' and
``lowering'' matrices, 
\begin{equation}
\sigma ^{+}=\frac{\sigma ^{x}+i\,\sigma ^{y}}{2}=\left( 
\begin{array}{cc}
0 & 1 \\ 
0 & 0
\end{array}
\right) \;\;\;\;\;\;\;\;\;\;{\rm and}\;\;\;\;\;\;\;\;\;\;\sigma ^{-}=\frac{%
\sigma ^{x}-i\,\sigma ^{y}}{2}=\left( 
\begin{array}{cc}
0 & 0 \\ 
1 & 0
\end{array}
\right) \,,  \label{flip}
\end{equation}
whose operations over the base vectors are $\sigma ^{+}\,e_{-}=e_{+}$, $%
\sigma ^{-}\,e_{+}=e_{-}$, and $\sigma ^{+}\,e_{+}=\sigma ^{-}\,e_{-}=0$.

The spin operator $\sigma _{i}^{\alpha }$ in (\ref{h}) acts on the Hilbert
space ${\cal H}$ and has the form 
\begin{equation}
\sigma _{i}^{\alpha }=I\otimes \cdots \otimes I\otimes \sigma ^{\alpha
}\otimes I\otimes \cdots \otimes I  \label{sigma}
\end{equation}
with $\sigma ^{\alpha }$ acting on ${\cal H}_{i}$ (i.e. it is located at $i$%
--th position). Here $\alpha $ runs over $x,y,z,0,+$ and $-$ with $\sigma
^{0}\equiv I$. The Hamiltonian is composed of pair interactions between
nearest neighbor sites, 
\[
\sigma _{i}^{\alpha }\,\sigma _{i+1}^{\beta }=I\otimes \cdots \otimes
I\otimes \sigma ^{\alpha }\otimes \sigma ^{\beta }\otimes I\otimes \cdots
\otimes I\,, 
\]
which acts as the identity operator in ${\cal H}_{k}$ for all $k$ except at
the positions $k=i,i+1$. The type of interactions considered in (\ref{h})
have pairs $(\alpha ,\beta )$ equal to $(x,x),(y,y)$ and $(z,z)$ which will
subsequently be written in terms of $(+,+),(+,-),(-,+)$ and $(-,-)$
interactions.

Note that two consecutive terms of the Hamiltonian (\ref{H}) do not commute, 
$\left[ h_{i-1,i},h_{i,i+1}\right] \not=0$, due to the fact that 
\[
\left[ \sigma _{i-1}^{\alpha }\,\sigma _{i}^{\beta },\sigma _{i}^{\gamma
}\,\sigma _{i+1}^{\delta }\right] =I\otimes \cdots \otimes \sigma ^{\alpha
}\otimes \left[ \sigma ^{\beta },\sigma ^{\gamma }\right] \otimes \sigma
^{\delta }\otimes I\otimes \cdots \otimes I 
\]
does not vanish if $\beta \not=\gamma $. This expresses the quantum
mechanical nature of our spin system. We shall now explain how Trotter's
formula resolves the non--commutativity character by converting quantum spin
variables $\{\sigma _{i}^{\alpha }\}$ into infinite many copies ${\bf s}_{1},%
{\bf s}_{2},\dots $ of ``classical'' spin variables ${\bf s}%
_{j}=\{s_{i,j}\}_{i=1}^{N}$.

For $N=2n$ let us decompose the set $\Lambda =\{1,2,\dots 2n\}=\Lambda ^{%
{\rm odd}}\cup \Lambda ^{{\rm even}}$, according to whether the site $i$ is
odd or even, and write 
\begin{equation}
H_{2n}=H_{n}^{{\rm odd}}+H_{n}^{{\rm even}}\,,  \label{HH}
\end{equation}
where $H_{n}^{{\rm odd}({\rm even})}$ is given by (\ref{H}) with $\Lambda $
replaced by $\Lambda ^{{\rm odd}({\rm even})}$. Because we have eliminated
consecutive pairings, the terms which compose $H_{n}^{{\rm odd}({\rm even})}$
commute with each other. Note, however, that $\left[ H_{n}^{{\rm odd}%
},H_{n}^{{\rm even}}\right] \not=0$.

Using Trotter's formula (\ref{Trotter}) we have 
\begin{equation}
\rho _{2n}=\lim_{m\rightarrow \infty }\left( e^{-\beta H_{n}^{{\rm odd}%
}/m}\;e^{-\beta H_{n}^{{\rm even}}/m}\right) ^{m}\equiv \lim_{m\rightarrow
\infty }\left( \rho _{n,m}^{{\rm odd}}\,\rho _{n,m}^{{\rm even}}\right)
^{m}\,.  \label{TF1}
\end{equation}
The partition function (\ref{Z}) can thus be written as 
\begin{equation}
Z_{2n}=\lim_{m\rightarrow \infty }Z_{n,m}  \label{Zn1}
\end{equation}
where 
\begin{equation}
Z_{n,m}={\rm Tr}\,\displaystyle\underbrace{\rho _{n,m}^{{\rm odd}}\,\rho
_{n,m}^{{\rm even}}\,\cdots \,\rho _{n,m}^{{\rm odd}}\,\rho _{n,m}^{{\rm even%
}}}_{m-{\rm times}}\,.  \label{Znm}
\end{equation}

We are going to make use of our base ${\cal B} $ by inserting the spectrum
decomposition of the identity (\ref{UM}) between all pairs of density
matrices $\rho ^{{\rm odd}} \,{\it I \! \! \! \! I} \, \rho ^{{\rm even}} $
and $\rho ^{{\rm even}} \, {\it I \! \! \! \! I} \, \rho ^{{\rm odd}} $. In
other words, we want to represent the density matrix $\rho $ according to
the base ${\cal B} $. To make this representation clearer we shall first
consider the special case $Z_{1,1}$.

Dropping, for simplicity, the subindex of $\rho $, we have 
\begin{equation}
Z_{1,1}={\rm Tr}\,\rho ^{{\rm odd}}\,\rho ^{{\rm even}}={\rm Tr}\,\rho ^{%
{\rm odd}}\,{\it I\!\!\!\!I}\,\rho ^{{\rm even}}\,{\it I\!\!\!\!I}=\sum_{%
{\bf s},{\bf s}^{\prime }}\rho _{{\bf s},{\bf s}^{\prime }}^{{\rm odd}%
}\,\rho _{{\bf s}^{\prime },{\bf s}}^{{\rm even}}  \label{TrA}
\end{equation}
where $\rho _{{\bf s},{\bf s}^{\prime }}=(e_{{\bf s}})^{T}\,\rho \,e_{{\bf s}%
^{\prime }}$. For $n=1$, $\Lambda =\{1,2\}$ has only two sites and ${\cal H}$
is a four dimensional space, ${\bf s}\in \{++,+-,-+,--\}$, leading to a $%
4\times 4$ density matrix $\rho $. In view of the periodic boundary
condition, $\rho ^{{\rm odd}}$ differs from $\rho ^{{\rm even}}$ by the
order of the spins at the sites $\{1,2\}$. We shall keep it in mind and drop
their super--index too. As we shall see, the alternating position of the
odd--even density matrix in the product (\ref{Znm}) will lead to a
chess-board pattern.

To compute the $4\times 4$ density matrix $\rho $, the algebraic properties
of the matrices (\ref{Pauli}) and (\ref{flip}) have to be used. Let us first
write $h_{1,2}$, with $J^{z}=0$, in terms of $\sigma ^{+},\sigma ^{-}$ and $%
\sigma ^{z}$. From (\ref{h}) and (\ref{flip}), we have 
\begin{equation}
h_{1,2}=(J^{x}-J^{y})\left( \sigma _{1}^{+}\,\sigma _{2}^{+}+\sigma
_{1}^{-}\,\sigma _{2}^{-}\right) +(J^{x}+J^{y})\left( \sigma
_{1}^{+}\,\sigma _{2}^{-}+\sigma _{1}^{-}\,\sigma _{2}^{+}\right)
+(h/2)\left( \sigma _{1}^{z}+\sigma _{2}^{z}\right) \,,  \label{h+-}
\end{equation}
where $J_{1,2}^{\alpha }=J_{2,1}^{\alpha }=J^{\alpha }$ and $h_{1}=h_{2}=h$.

Writing $\eta =(J^{x}-J^{y})$, $\kappa =(J^{x}+J^{y})$ and $\zeta =\sqrt{%
\eta ^{2}+h^{2}}$, the following relations

\begin{equation}
(h_{1,2})^{2k}\,e_{{\bf s}}=\left\{ 
\begin{array}{lll}
\zeta ^{2k}\,e_{++}\,, & {\rm if} & {\bf s}=++\,, \\ 
\kappa ^{2k}\,e_{+-}\,, & {\rm if} & {\bf s}=+-\,, \\ 
\kappa ^{2k}\,e_{-+}\,, & {\rm if} & {\bf s}=-+\,, \\ 
\zeta ^{2k}\,e_{--}\,, & {\rm if} & {\bf s}=--\,,
\end{array}
\right.  \label{rel1}
\end{equation}
and 
\begin{equation}
(h_{1,2})^{2k-1}\,e_{{\bf s}}=\left\{ 
\begin{array}{lll}
\zeta ^{2k-2}\left\{ h\,e_{++}+\eta \,e_{--}\right\} \,, & {\rm if} & {\bf s}%
=++\,, \\ 
\kappa ^{2k-1}\,e_{-+}\,, & {\rm if} & {\bf s}=+-\,, \\ 
\kappa ^{2k-1}\,e_{+-}\,, & {\rm if} & {\bf s}=-+\,, \\ 
\zeta ^{2k-2}\left\{ -h\,e_{--}+\eta \,e_{++}\right\} \,, & {\rm if} & {\bf s%
}=--\,,
\end{array}
\right.  \label{rel2}
\end{equation}
hold for any integer number $k\geq 1$.

Taylor--expanding $\rho =e^{\beta \,h_{1,2}}$ and using equations (\ref{rel1}%
) and (\ref{rel2}), combined with the orthogonal relation $(e_{{\bf s}%
})^{T}\,e_{{\bf s}^{\prime }}=\delta _{{\bf s},{\bf s}^{\prime }}$, gives 
\begin{equation}
\rho =\left( 
\begin{array}{cccc}
\cosh \beta \zeta +(h/\zeta )\sinh \beta \zeta & 0 & 0 & (\eta /\zeta )\sinh
\beta \zeta \\ 
0 & \cosh \beta \kappa & \sinh \beta \kappa & 0 \\ 
0 & \sinh \beta \kappa & \cosh \beta \kappa & 0 \\ 
(\eta /\zeta )\sinh \beta \zeta & 0 & 0 & \cosh \beta \zeta -(h/\zeta )\sinh
\beta \zeta
\end{array}
\right) \,.  \label{rhoM}
\end{equation}
Note that $\rho $ is a symmetric matrix and $Z_{1,1}={\rm Tr}\,\rho ^{2}$.

Now, we want to include the $J^{z}$ term into the above calculation. The
simplest way to deal with this case is to introduce a slightly modification
of the Trotter formula \cite{KM}, 
\[
e^{K_{0}+K_{1}+K_{2}}=\lim_{m\rightarrow \infty }\left(
e^{K_{0}/m}\,e^{K_{1}/m}\,e^{K_{2}/m}\right) ^{m}\,. 
\]
As in \cite{BS} we separate the $J^{z}$--dependent terms of $H_{2n}$ and
divide the remaining terms according to the even--odd partition of $\Lambda $%
, 
\begin{equation}
H_{2n}=H_{2n}^{0}+{H}_{n}^{{\rm odd}}+{H}_{n}^{{\rm even}}\,,  \label{0eo}
\end{equation}
where ${H}_{n}^{{\rm odd}({\rm even})}$ is as in (\ref{HH}). We repeat the
steps (\ref{TF1}) -- (\ref{Znm}) with the modified Trotter formula.

For the special case $Z_{1,1}$, we have 
\begin{equation}
Z_{1,1}={\rm Tr}\,\rho ^{0}\,{\rho }^{{\rm odd}}\,{\rho }^{{\rm even}}={\rm %
Tr}\,\left( \rho ^{0}\right) ^{1/2}{\rho }^{{\rm odd}}\,{\rho }^{{\rm even}%
}\left( \rho ^{0}\right) ^{1/2}=\sum_{s,s^{\prime }}\,\hat{\rho}%
_{s,s^{\prime }}^{{\rm odd}}\,\hat{\rho}_{s^{\prime },s}^{{\rm even}}\,,
\label{Z11}
\end{equation}
where $\hat{\rho}^{{\rm odd}}=\left( \rho ^{0}\right) ^{1/2}\,\rho \equiv 
\hat{\rho}$ with $\rho $ given by (\ref{rhoM}), $\hat{\rho}^{{\rm even}%
}=\left( \hat{\rho}\right) ^{T}$ and 
\begin{equation}
\rho ^{0}=e^{2\beta J^{z}\,\sigma ^{z}\otimes \sigma ^{z}}=\left( 
\begin{array}{cccc}
e^{2\beta J^{z}} & 0 & 0 & 0 \\ 
0 & e^{-2\beta J^{z}} & 0 & 0 \\ 
0 & 0 & e^{-2\beta J^{z}} & 0 \\ 
0 & 0 & 0 & e^{2\beta J^{z}}
\end{array}
\right) \,.  \label{rho0}
\end{equation}
Note that $\hat{\rho}=(\hat{\rho})^{T}$.


\medskip

To extend the calculation of $Z_{1,1}$ to $Z_{n,m}$ with $n,m\geq 1$, we may
think of a rectangular $2m\times 2n$ lattice as a ``chess--board'' with $%
n\cdot m$ black squares. To each of these black pieces labeled by $p$ (of
plaquetes) we associate a matrix $\rho _{p}$ given by (\ref{rhoM}) with $%
\beta $ replaced by $\beta /m$; $\rho $ is a transfer matrix between the
state of spins at the up corners of the black square $p$ and the state of
the spins at the bottom corners of $p$. If the up and bottom corners' labels
of $p$ are respectively $v,w$ and $v^{\prime },w^{\prime }$, we denote the
matrix elements of $\rho _{p}$ by $\rho (s_{v},s_{w};s_{v^{\prime
}},s_{w^{\prime }})$ and equation (\ref{Znm}) represented in the base ${\cal %
B}$ can be written as 
\begin{equation}
Z_{n,m}=\sum_{\{s_{i,j}\}}\prod_{j=1}^{m}\left( \prod_{i=1}^{n}\rho
(s_{2i-1,2j-1},s_{2i,2j-1};s_{2i-1,2j},s_{2i,2j})\,\rho
(s_{2i,2j},s_{2i+1,2j};s_{2i,2j+1},s_{2i+1,2j+1})\right) \,,  \label{prodrho}
\end{equation}
with periodic conditions assumed in both directions. For (\ref{prodrho}),
note that the matrix elements of $\rho ^{{\rm odd}({\rm even})}$ between two
successive spin configurations, ${\bf s}_{l}=\{s_{k,l}\}_{k=1}^{2n}$ and $%
{\bf s}_{l+1}=\{s_{k,l+1}\}_{k=1}^{2n}$, factor out 
\[
\left( \rho _{n,m}^{{\rm odd}}\right) _{{\bf s}_{2j-1},{\bf s}%
_{2j}}=\prod_{i=1}^{n}\rho
(s_{2i-1,2j-1},s_{2i,2j-1};s_{2i-1,2j},s_{2i,2j})\,, 
\]
and analogously for $\rho _{n,m}^{{\rm even}}$, due to the even--odd
decomposition.

As a consequence, two matrices $\rho _{p}$ and $\rho _{p^{\prime }}$ are
coupled if $p$ and $p^{\prime }$ share the same vertex $v$ ($p\cap p^{\prime
}=v$). Each matrix labeled by a plaquete $p$ has a pairwise interaction with
four other matrices labeled by the nearest--neighbor black plaquetes of $p$.
We shall use this fact in the following section to write (\ref{prodrho}) as
a counting problem in a $n\times m$ rectangular lattice (wrapped into a
torus and rotated $45^{\circ }$) whose vertices are centers of the black
plaquetes of the original lattice.


\section{The Vertex Model}

\label{vertex} \setcounter{equation}{0} \setcounter{theorem}{0}

In the previous section Trotter's formula has been used to write the
partition function $Z_{2n}$ of a one--dimensional quantum spin system in $%
\Lambda _{2n}=\{1,\dots ,2n\}$ as the $m\rightarrow \infty $ limit of a
sequence $\{Z_{n,m}\}_{m\geq 1}$ of partition functions of two--dimensional
classical spin systems in $\Lambda _{n,m}\equiv \Lambda _{2n}\times \Lambda
_{2m}$ with four spin interactions mediated by the matrix $\rho $ (see
equations (\ref{Zn1}), (\ref{Znm}) and (\ref{prodrho})). The mapping we
shall now discuss is due to Barma and Shastry \cite{BS}.

We consider $\Lambda _{n,m}$ wrapped in a torus as a ``chess-board'' and
define a $n\times m$ regular lattice $\Lambda _{n,m}^{\ast }$ with vertices
at the center of the black plaquetes of $\Lambda _{n,m}$ and edges
connecting nearest neighbor vertices (see Figure $1$). Note that the edges
of $\Lambda _{n,m}^{\ast }$ are $45^{\circ }$ with respect to edges of $%
\Lambda _{n,m}$. The remarkable property in this construction is the {\it %
one--to--one correspondence} between vertices of $\Lambda _{n,m}$ and edges
of $\Lambda _{n,m}^{\ast }$.

\begin{figure}[ht]
  \centering
  \includegraphics[width=1.7417in,height=1.7279in]{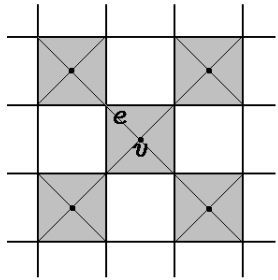}
  \caption{The regular $\Lambda_{n,m}$ and dual $\Lambda _{n,m}^{\ast }$ lattices}
\end{figure}

Let us define the ``dual'' relation $\ast :\Lambda _{n,m}\longrightarrow
\Lambda _{n,m}^{\ast }$, defined by 
\[
(i,j)^{\ast }=e\,, 
\]
if the the edge $e\in \Lambda _{n,m}^{\ast }$ crosses the site with
coordinates $(i,j)\in \Lambda _{n,m}$, and
\[
p^{\ast }=v\,, 
\]
if the vertex $v\in \Lambda _{n,m}^{\ast }$ is the center of a black
plaquete $p\in \Lambda _{n,m}$. This allows us to establish a correspondence
between classical spin configurations $\{s_{i,j}:i=1,\dots ,2n;j=1,\dots
,2m\}$ and the two--state edge configurations in $\Lambda _{n,m}^{\ast }$.
We say that the edge $e=(i,j)^{\ast }$ is occupied or vacant according to
whether the value of $s_{i,j}$ is $1$ or $-1$.

Now, the density matrix $\rho $ at the plaquete $p$, whose values depend on
the spin configuration of its four corners, is mapped, by the ``dual''
transformation, into a vertex function $w$ at $v=p^{\ast }$ whose values
depend on the configuration $\xi _{v}$ of the four incident edges. In the
period $1$ system there is only one vertex function $w$ for all $v\in
\Lambda _{n,m}^{\ast }$. This leads us to the following definitions.

A {\it vertex configuration} $\xi $ is an assignment $\xi :\Lambda
_{n,m}^{\ast }\longrightarrow \{1,\dots ,8\}$ which associates to each
vertex $v$ one of the $4$--incident--edges configurations $\xi _{v}$ shown
in Figure 2. There are actually $2^{4}=16$ possible assignments of
occupied--vacant incident edges but only $8$ of them (shown in Figure 2) are
relevant to the problem at hand.

\begin{figure}[ht]
  \centering      
  \includegraphics[width=2.7804in,height=1.9104in]{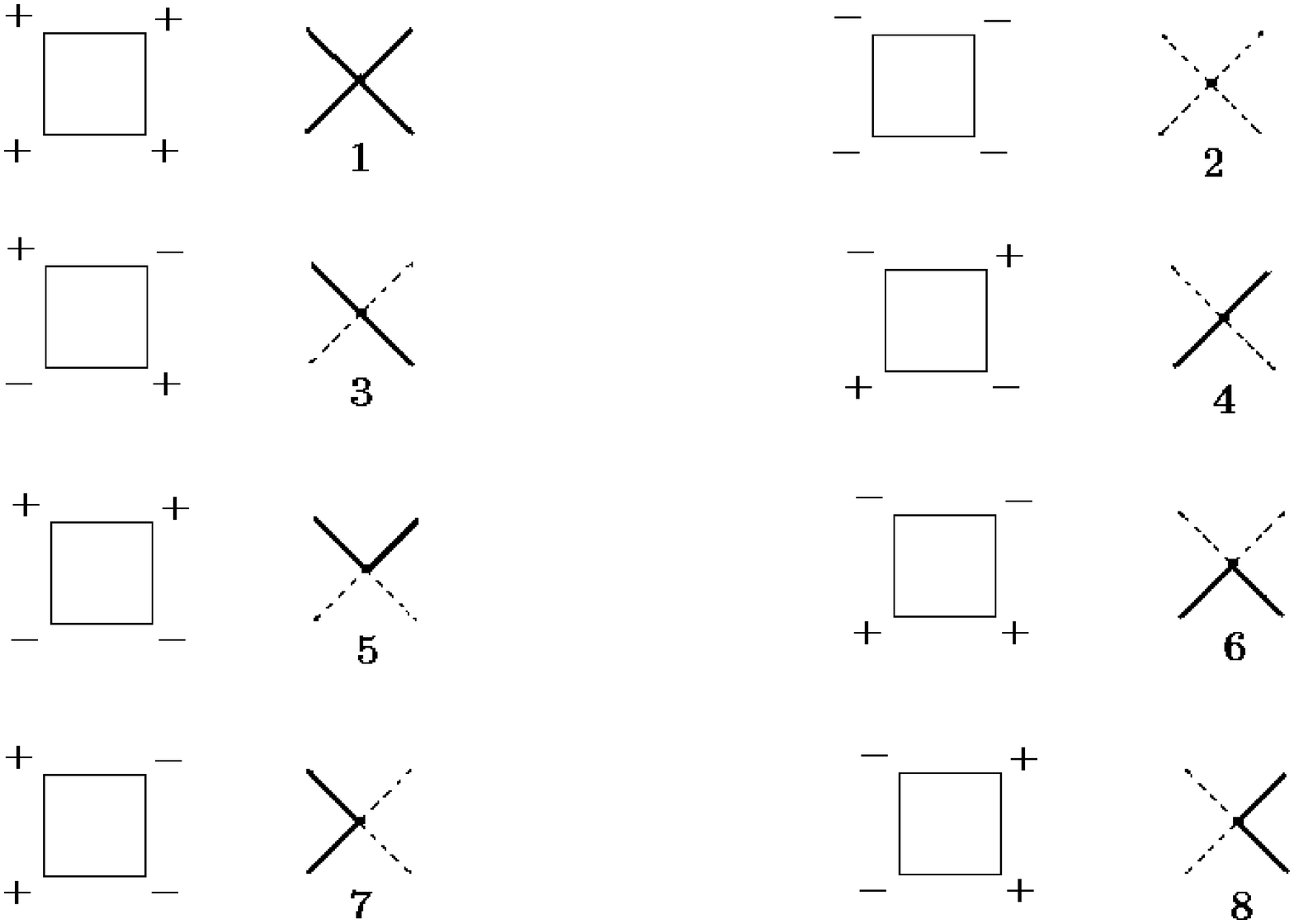}
  \caption{Possible assignements  of $\protect\xi _{v}$}
\end{figure}

A vertex configuration $\xi $ is said to be {\it compatible} if for each
edge $e$ with end points $v$ and $v^{\prime }$, $\xi _{v}$ and $\xi
_{v^{\prime }}$ have the same assignment for the common edge $e$. From here
on, by vertex configuration $\xi $ we always means a compatible
configuration. It is worth noting that a vertex configuration $\xi $
corresponds to a covering of $\Lambda _{n,m}^{\ast }$ by closed polygons
(see Figure 3).

\begin{figure}[ht]
  \centering      
  \includegraphics[width=2.1413in,height=1.8922in]{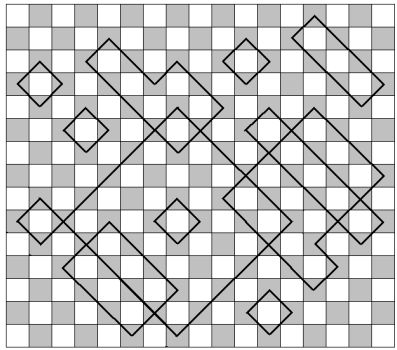}
  \caption{A vertex configuration $ \protect\xi $}
\end{figure}

The weight of a vertex configuration $\xi $ is given by 
\begin{equation}
w(\xi )=\prod_{v\in \Lambda _{n,m}^{\ast }}w_{\xi _{v}}\,,  \label{wxi}
\end{equation}
where, in view of (\ref{rhoM}), (\ref{Z11}), the ``dual'' mapping and the
label chosen in Figure 2, 
\begin{equation}
\begin{array}{lllll}
&  & w_{1} & = & e^{\beta J^{z}/m}\left( \cosh \displaystyle\frac{\beta
\zeta }{m}+\displaystyle\frac{h}{\zeta }\,\sinh \displaystyle\frac{\beta
\zeta }{m}\right) \,, \\ 
&  & w_{2} & = & e^{\beta J^{z}/m}\left( \cosh \displaystyle\frac{\beta
\zeta }{m}-\displaystyle\frac{h}{\zeta }\,\sinh \displaystyle\frac{\beta
\zeta }{m}\right) \,, \\ 
w_{3} & = & w_{4} & = & e^{-\beta J^{z}/m}\,\sinh \displaystyle\frac{\beta
\kappa }{m}\,, \\ 
w_{5} & = & w_{6} & = & e^{-\beta J^{z}/m}\,\cosh \displaystyle\frac{\beta
\kappa }{m}\,, \\ 
w_{7} & = & w_{8} & = & e^{-\beta J^{z}/m}\displaystyle\frac{\eta }{\zeta }%
\,\sinh \displaystyle\frac{\beta \zeta }{m}\,,
\end{array}
\label{w}
\end{equation}
where $\zeta $, $\eta $ and $\kappa $ are given following equation (\ref{h+-}%
). Note that only the eight vertex configurations shown in Figure 2 have
non--vanishing vertex function.

With these definitions, we conclude

\begin{proposition}
\label{Vertex} The partition function (\ref{prodrho}) can be written as 
\begin{equation}
Z_{n,m}=\sum_{\xi }w(\xi )\,,  \label{iden}
\end{equation}
where the sum runs over all compatible vertex configurations $\xi $ with
weights given by (\ref{wxi}) and (\ref{w}).
\end{proposition}

At this point, one can verify whether (\ref{w}) satisfies the {\it free
fermion condition} stated in (\ref{ff}). The equation

\[
w_{1}w_{2}+w_{3}w_{4}-w_{5}w_{6}-w_{7}w_{8}= \left( e^{2\beta
J^{z}/m}-e^{-2\beta J^{z}/m} \right) \left( 1+ \frac{\eta ^2}{\zeta ^2} \sinh
^2 \frac{\beta \zeta}{m}\right) 
\]
vanishes if and only if $J^{z}=0$. This will be assumed in the next sections.


\section{Pfaffian Solution}

\label{pfaffian} \setcounter{equation}{0} \setcounter{theorem}{0}

In this section the partition function $Z_{n,m}$, written as the generating
function of a spanning subgraph of $\Lambda ^{\ast }\equiv \Lambda
_{n,m}^{\ast }$ in (\ref{iden}), will be solved by the Pfaffian technique.
For this, we introduce some terminology.

Let ${\cal G}$ be a graph which we may think to be a ``decoration'' of $%
\Lambda ^{\ast }$. A {\it dimer covering} $D$ of ${\cal G}$ is a spanning
subgraph consisting of non--overlapping close--packed configurations of
elements formed by an edge and its two terminal vertices (dimer). Note that
no vertex is left vacant in $D$.

In the early $60$'s the problem of enumerating all dimer coverings has been
solved for certain lattice graphs (see \cite{K} for a review). Right after
this, Green and Hurst \cite{GH} showed that the counting dimer coverings can
be useful to solve the partition function of two--dimensional Ising models
(see also \cite{GH}). Since the Ising model has to be written as the
generating function of a spanning subgraph of the form (\ref{iden}), the
method we now start to explain can be applied directly to this equation.

Given a collection of positive numbers ${\bf z}=\{z_{e}\}_{e\in {\cal G}}$
indexed by the edges of ${\cal G}$, let $\Gamma ({\bf z})$ be the generating
function for the dimer covering problem on ${\cal G}$, 
\begin{equation}
\Gamma ({\bf z})=\sum_{D}z({D})\,,  \label{gfcov}
\end{equation}
where 
\[
z({D})=\prod_{e\in D}z_{e} 
\]
is the weight of a dimer covering $D$ with the product running over edges
covered by the dimers of $D$.

If $\{1,2,\dots ,k\}$ is an arbitrary enumeration of the vertices of ${\cal G%
}$ let $A=\{a_{xy}\}$ be a skew--symmetric matrix of order $k$ whose
elements $a_{xy}=-a_{yx}$ satisfy 
\begin{equation}
|a_{xy}|=\left\{ 
\begin{array}{lll}
z_{e}\,, &  & {\rm if}\;\,e=\langle xy\rangle \in {\cal G}\,, \\ 
&  &  \\ 
0\,, &  & {\rm otherwise}\,.
\end{array}
\right.  \label{m}
\end{equation}
The sign of $a_{xy}$ is chosen according to a given orientation of the edges
of ${\cal G}$: $a_{xy}>0$ if $e$ is oriented from $x$ to $y$.

Our ability to calculate the partition function (\ref{iden}) depends \ \ \ \
\ \ \ on whether the graph ${\cal G}$ obtained by decorating $\Lambda ^{\ast
}$, and the associate weight function, agree with the hypotheses of the
following results.

\begin{proposition}
\label{det} Let $A$ be a $k\times k$ matrix defined as above with $k$ even.
Then 
\begin{equation}
\det A=({\rm Pf}\,A)^{2}\,,  \label{det-pf2}
\end{equation}
where the Pfaffian, ${\rm Pf}\,A$, of $A$ is defined by 
\[
{\rm Pf}\,A=\sum_{\pi }\varepsilon _{P_{\pi
}}\,a_{j_{1},j_{2}}\,a_{j_{3},j_{4}}\,\cdots \,a_{j_{k-1},j_{k}}\,, 
\]
with the sum over all partitions of $\{1,2,\dots ,k\}$ into pairs; $P_{\pi }$
is a permutation $j_{1}\,j_{2}\,\cdots \,j_{k}$ such that $%
|j_{1}\,j_{2}|j_{3}\,j_{3}|\cdots |j_{k-1}\,j_{k}|$ is a description of the
partition $\pi $ and $\varepsilon _{P_{\pi }}$ is the signature of $P_{\pi }$
\end{proposition}

For a proof of Proposition \ref{det} see \cite{K}. To understand, however,
equation (\ref{det-pf2}) and get a feeling of the next statement, let us
observe that if we {\it superpose} a dimer covering $D$ {\it on} another
dimer covering $D^{\prime }$ we obtain a {\it circuit covering} of ${\cal G}$%
. For, starting from any vertex of ${\cal G}$ we walk along the edge in $D$
till we reach other extremity; then, walk along the edge of $D^{\prime }$
and along the edge of $D$ after the extremity has been reached and so on.
Continuing this process we eventually arrive at the starting vertex when we
have to choose a different one and start again this process till all
vertices of ${\cal G}$ have been covered. It is a matter of fact that $\det
A $ is the generating function of circuit covering of ${\cal G}$. Note that 
\[
\det A=\sum_{P}\varepsilon _{P}\,a_{1,P(1)}\cdots a_{k,P(k)}\,, 
\]
and all permutation $P=\{P(1),\dots ,P(k)\}$ of $\{1,\dots ,k\}$ can be
decomposed into cycles $(C_{1},\dots ,C_{s})$. Since by the superposition of 
$D$ on $D^{\prime }$ with $D,D^{\prime }$ running over all dimer covering
one gets all circuit covering of ${\cal G}$, it is tempting to identify $%
\det A$ with $\Gamma ^{2}(z)$. As we shall see, this requires to choose the
orientation of ${\cal G}$ properly.

\begin{proposition}
\label{pfaff} If the number of contra--oriented steps, with respect to the
chosen orientation of ${\cal G}$, performed along each circuit generated
from the superposition of $D$ on $D^{\prime }$ is odd, for any two dimer
coverings $D$ and $D^{\prime }$ of ${\cal G}$, then 
\begin{equation}
\Gamma ({\bf z})=|{\rm Pf}\,A|\,.  \label{Gammapf}
\end{equation}
\end{proposition}

By imposing that the parity of all circuits to be odd $\det A$ becomes a sum
of positive terms\footnote{%
Note that the signature $\varepsilon _{P}$ of a permutation $P$ is given by $%
(-1)^{s}$ where $s$ is the number of cycles $(C_{1},\dots ,C_{s})$ in $P$.
If the parity of all circuits is odd, we get $(-1)$ from the product of
matrix elements along each cycle $C_{i}$ compensating the signature.} and
consequently the sign of every term in Pf$\,A$ remains the same implying
Proposition \ref{pfaff} (see ref. \cite{K}, pg. $89$, for a proof). We shall
call an orientation of ${\cal G}$ with such property {\it admissible}.

Combining the two propositions we arrive at 
\begin{equation}
\Gamma ({\bf z})=\sqrt{\det A}\,.  \label{Gammadet}
\end{equation}
It thus remains only to write (\ref{iden}) as a generating function of dimer
covering of a properly oriented graph ${\cal G}$ and evaluate the
determinant of the associated skew-symmetric matrix $A$.

\medskip

We begin by choosing the appropriated ``decorated'' graph ${\cal G}$. We
take ${\cal G}$ as being the {\it terminal graph} $(\Lambda ^{\ast })^{T}$
(see Figure $4$) obtained by the replacement of all vertices $v\in \Lambda
^{\ast }$ by a cluster of $4$ vertices all connected with each other by
single edges (a complete graph $K_{4}$) denoted by $\Box _{v}$. Each vertex
of ${\cal G}$ is an {\it extreme} or {\it terminal} of an incident edge $%
e\in \Lambda ^{\ast }$ a property by which ${\cal G}$ is named. The edge $%
e\in {\cal G}$ will be called {\it internal} if it belongs to $\Box _{v}$
for some $v$ and \ \ \ {\it external} otherwise. The external edges are
identified with the edges of $\Lambda ^{\ast }$. Note that the graph ${\cal G%
}$ is not planar due to the crossing of diagonal edges of $\Box _{v}$ which
prevents, in general, that an admissible orientation can be found leading
the Pfaffian method useless to compute $\Gamma ({\bf z})$. As we shall see,
it is exactly this problem that will help us to solve another one.

\begin{figure}[ht]    
  \centering      
  \includegraphics[width=2.9404in,height=1.5532in]{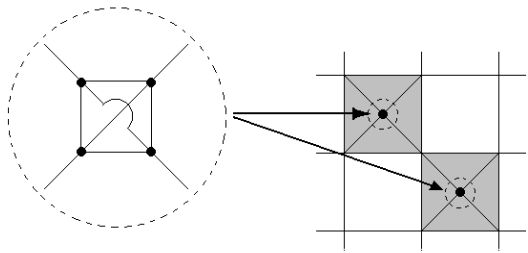}
  \caption{Construction of \ the  terminal graph $(\Lambda ^{\ast })^{T}$ }
\end{figure}

What makes a terminal graph significant for our problem is that we can
associate to each (compatible) vertex configuration $\xi $ in $\Lambda
^{\ast }$ a dimer covering $D$ in the terminal lattice ${\cal G}$ as
follows: an external edge $e\in D$ if the edge $e\in \Lambda ^{\ast }$ is
occupied in the configuration $\xi $. A list of vertex configurations $\xi
_{v}$ at vertex $v$ and its associated dimer covering at $\Box _{v}$ is
shown in Figure 5. The problem here is the degeneracy of dimer covering
associated with $\xi _{v}=2$.

\begin{figure}[ht]    
  \centering      
  \includegraphics[width=3.7075in,height=2.1171in]{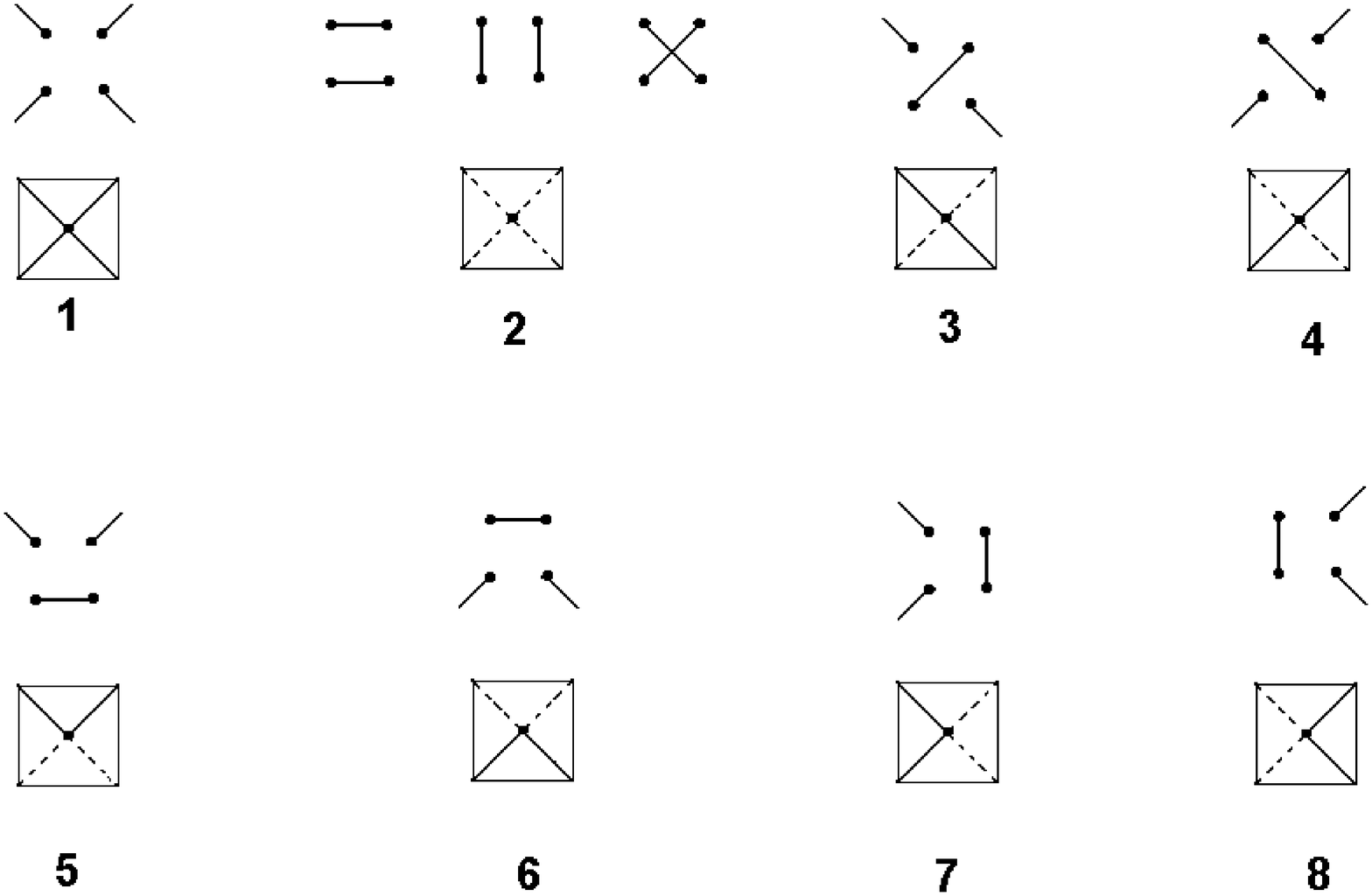}
  \caption{Vertex and their  associated dimer configurations}
\end{figure}

It turns out that there exist a (non--admissible) orientation of ${\cal G}$
such that, if a dimer configuration $D^{\prime }$ differ from $D$ by the
substitution of a crossing dimer covering at $\Box _{v}$ by a non-crossing
one, the superposition of $D^{\prime }$ on $D$ yields a circuit in $\Box
_{v} $ with even parity and a minus sign is given to this contribution. In
addition, this orientation is such that all circuits, excluding the ones
with crossings and those winding around the torus, generated by the
superposition of $D$ on $D^{\prime }$, with $D$ and $D^{\prime }$ two dimer
coverings in ${\cal G}$, have odd parity. The orientation of ${\cal G}$ with
these properties is shown in Figure $6$ for a given $\Box _{v}$.

\begin{figure}[ht]    
  \centering      
  \includegraphics[width=1.6648in,height=1.6302in]{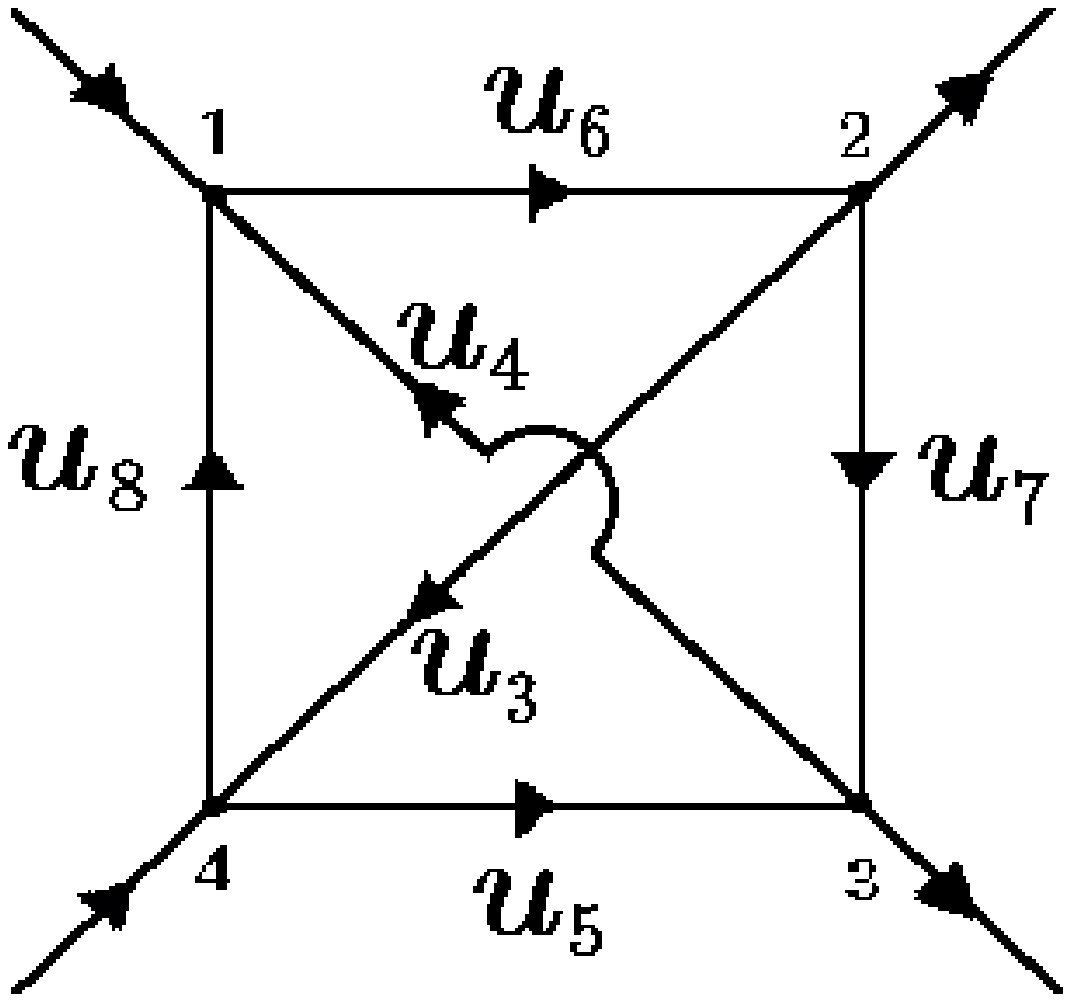}
  \caption{The orientation of a cell $\Box _{v}$}
\end{figure}

As demonstrated in ref. \cite{K}, pages $98$--$99$, it is the crossing dimer
configuration which gives rise to the minus sign and Proposition \ref{pfaff}
holds if $z(D)$ is replaced by ${\tilde{z}}(D)=(-1)^{c}\,z(D)$ in equation (%
\ref{gfcov}) where $c$ is the number of $v$'s such that $\Box _{v}$ is
cross--covered in $D$ and free boundary condition is imposed in ${\cal G}$.

As $\Lambda _{n,m}$ is wrapped on a torus, $\Lambda _{n,m}^{\ast }$ may have
many different topologies depending on the values of $(n,m)$. The lattice $%
\Lambda _{n,m}^{\ast }$ will be wrapped on a torus only if $m$ is a multiple
of $n$. We shall in the following assume that $m=kn+1$ for some integer $k$
which makes $\Lambda _{n,m}^{\ast }$ to be {\it helically wound} on a torus
and avoids circuits winding around ${\cal G}=(\Lambda _{n,m}^{\ast })^{T}$
(see McCoy and Wu \cite{McW} for the dimer covering problem on a torus).

\medskip

Once the orientation of ${\cal G} $ has been resolved, we are in position to
write (\ref{iden}) as a generating function $\Gamma ({\bf z})$ of dimer
covering of ${\cal G} = (\Lambda ^{\ast})^{T}$. We observe that the
degeneracy in the association of $\xi _{v}=2$ results from the dimer
covering of $\Box _{v}$ using only internal edges. For all dimer covering
associated with $\xi _{v}\not =2$ the association is unique and there are at
least two external edges covered.

Now, let ${\cal C}_{\xi }$ denote the class of equivalence defined as
follows. Two dimer covering of ${\cal G}$, $D$ and $D^{\prime }$, are said
to be equivalent, $D\sim D^{\prime }$, if they give rise a configuration $%
\xi $ under the association made in Figure $5$. Then equation (\ref{gfcov})
can be written as 
\begin{equation}
\Gamma ({\bf z})=\sum_{\xi }u({\xi })\,,  \label{gfcov1}
\end{equation}
where 
\begin{equation}
u({\xi })=\sum_{D\in {\cal C}_{\xi }}z(D)\,.  \label{zxi}
\end{equation}

As Fan and Wu \cite{FW} (see also \cite{HLW}), if in addition we set $%
z_{e}=1 $ for all external edges $e\in {\cal G}$, then $u(\xi )$ factors
out, 
\begin{equation}
u({\xi })=\prod_{v\in \Lambda ^{\ast }}u_{\xi _{v}}\,,  \label{z-xi}
\end{equation}
\nolinebreak\ where $u_{\xi _{v}}$ depends only on $\{z_{e}\}_{e\in \Box
_{v}}$ and is explicitly given by (using the label of Figure $5$) 
\begin{equation}
u_{\xi }=\left\{ 
\begin{array}{lll}
1\,, & {\rm if} & \xi =1\,, \\ 
z_{\langle 14\rangle }\,z_{\langle 23\rangle }+z_{\langle 34\rangle
}\,z_{\langle 12\rangle }-z_{\langle 13\rangle }z_{\langle 24\rangle }\,, & 
{\rm if} & \xi =2\,, \\ 
z_{\langle 24\rangle }\,, & {\rm if} & \xi =3\,, \\ 
z_{\langle 13\rangle }\,, & {\rm if} & \xi =4\,, \\ 
z_{\langle 34\rangle }\,, & {\rm if} & \xi =5\,, \\ 
z_{\langle 12\rangle }\,, & {\rm if} & \xi =6\,, \\ 
z_{\langle 23\rangle }\,, & {\rm if} & \xi =7\,, \\ 
z_{\langle 14\rangle }\,, & {\rm if} & \xi =8\,.
\end{array}
\right.  \label{zweight}
\end{equation}

In order to compare equation (\ref{iden}) with (\ref{gfcov1}), we identify $%
u(\xi )$ with the normalized vertex function $w(\xi )/w_{1}$: $u_{\xi
_{v}}=w_{\xi _{v}}/w_{1}$. Equation (\ref{Vertex}) then reads 
\[
Z_{n,m}=w_{1}^{n\cdot m}\sum_{\xi }u(\xi )=w_{1}^{n\cdot m}\cdot \Gamma (%
{\bf z})\,. 
\]
Note that ${\bf z}=\{z_{e}\}$ has to be specified only for the six internal
edges of $\Box _{v}$ for a given $v$ (recall $z_{e}=1$ for all external
edges). Moreover, note that the last six equations of (\ref{zweight})
determine completely the correspondence between the vertex function $u(\xi
)=w(\xi )/w_{1}$ and the weight ${\bf z}$ since, by substituting those
equations into the second equation, we have 
\begin{equation}
u_{2}=u_{5}\,u_{6}+u_{7}\,u_{8}-u_{3}\,u_{4}=\frac{w_{5}\,w_{6}+w_{7}%
\,w_{8}-w_{3}\,w_{4}}{w_{1}^{2}}=\frac{w_{2}}{w_{1}}\,,  \label{u-zz}
\end{equation}
provided $w_{1},\dots ,w_{8}$ satisfy the free--fermion condition (\ref{ff}).

In view of (\ref{Gammadet}), the conclusion of this section may be summarize
by the following result.

\begin{proposition}
\label{Z=det} Let $A$ be a $(4nm\times 4nm)$ skew--symmetric matrix defined
as in (\ref{m}) with the sign determined by the orientation of ${\cal G}$ as
in Figure $6$. If $z_{e}=1$ for all external edges and the equations $u_{\xi
}=w_{\xi }/w_{1}$, $\xi \in \{3,4,\dots ,8\}$, given by (\ref{zweight})
define $z_{e}$ for all internal edges $e\in {\cal G}$, then 
\begin{equation}
Z_{n,m}=w_{1}^{n\cdot m}\,\sqrt{\det A}\,.  \label{Z-det}
\end{equation}
\end{proposition}


\section{Evaluation of the Determinant}

\label{Det} \setcounter{equation}{0} \setcounter{theorem}{0}

In this section we evaluate the determinant of the $4nm\times 4nm$ matrix $A$%
, with entries given by (\ref{m}), as described in Proposition \ref{Z=det}.

We begin by block--diagonalizing $A$ by the Fourier method. Let us first
remind some facts about Kronecker (or tensor) product. If $M=\{m_{ij}\}$ and 
$N=\{n_{ij}\}$ are $k\times k$ and $l\times l$ matrices, respectively, the 
{\it Kronecker product} of $M$ and $N$, $M\otimes N$, is a $kl\times kl$
matrix given by $M\otimes N=\{m_{ij}N\}$. If $M_{1}\otimes N_{1}$ and $%
M_{2}\otimes N_{2}$ are two matrices as above, their product is given by 
\begin{equation}
\left( M_{1}\otimes N_{1}\right) \,\left( M_{2}\otimes N_{2}\right)
=M_{1}\,M_{2}\otimes N_{1}\,N_{2}\,.  \label{mrule}
\end{equation}
The Kronecker products do not commute, $M\otimes N\neq N\otimes M$, but
there exists a permutation matrix $P$ such that $P^{T}\left( M\otimes
N\right) P=N\otimes M$. So 
\[
\det M\otimes N=\det N\otimes M=\det \left( I\otimes M\right) \,\det \left(
I\otimes N\right) =\left( \det M\right) ^{l}\,\left( \det N\right) ^{k}\,. 
\]

We want to write $A$ as a $nm\times nm$ matrix whose elements are $4\times 4$
matrices, called inner matrices. Since the entry $a_{xy}$ of $A$ is
nonvanishing if $\langle xy\rangle =e$ is either an edge at $\Box _{v}$ or
connecting nearest neighbor ones, there are only three relevant inner
matrices. The inner matrices are labeled by the enumeration of vertices of $%
\Box _{v}$ and the sign of their entries determined according to the
orientation of ${\cal G}$ (see Figure $6$).

Let $U$ be the skew-symmetric ``adjacency matrix'' of $\Box _{v}$ defined by
the last six equations of (\ref{zweight}), 
\begin{equation}
U=\left( 
\begin{array}{cccc}
0 & u_{6} & -u_{4} & -u_{8} \\ 
-u_{6} & 0 & u_{7} & u_{3} \\ 
u_{4} & -u_{7} & 0 & -u_{5} \\ 
u_{8} & -u_{3} & u_{5} & 0
\end{array}
\right) \,,  \label{U}
\end{equation}
and let $X$ and $Y$ be defined by the external edges bridging $\Box _{v}$
and $\Box _{v^{\prime }}$ in the southeast and southwest directions,
respectively, 
\begin{equation}
X=\left( 
\begin{array}{cccc}
0 & 0 & 0 & 0 \\ 
0 & 0 & 0 & 0 \\ 
1 & 0 & 0 & 0 \\ 
0 & 0 & 0 & 0
\end{array}
\right) \;\;\;\;\;\;\;\;\;\;\;{\rm and}\;\;\;\;\;\;\;\;\;\;\;Y=\left( 
\begin{array}{cccc}
0 & 0 & 0 & 0 \\ 
0 & 0 & 0 & 1 \\ 
0 & 0 & 0 & 0 \\ 
0 & 0 & 0 & 0
\end{array}
\right) \,.  \label{XY}
\end{equation}

The $nm\times nm$ matrix is labeled by the vertices $v$ of $\Lambda
_{n,m}^{\ast }$ which is helically wound on the torus (recall $m=kn+1$). The
vertex enumeration $\{1,2,\dots ,mn\}$ can be made starting from the upper
left corner following the southeast direction. Note that all vertices are
visited in this way with the last site $mn$ being neighbor to the first one.
In addition, the four nearest neighbors of a vertex $j$ have the following
coordinates (modulo $mn$): in the northeast and southwest directions, $j^{%
{\rm ne}}=j+m$ and $j^{{\rm sw}}=j-m$, respectively; in the northwest and
southeast neighbors, we have $j^{{\rm nw}}=j-1$ and $j^{{\rm se}}=j+1$.

Here, the relevant $N\times N$ matrices are the identity $I=I_{N}$ and the
``forward shift'' permutation, 
\begin{equation}
\Pi =\Pi _{N}=\left( 
\begin{array}{ccccc}
0 & 1 & 0 & \cdots & 0 \\ 
0 & 0 & 1 & \cdots & 0 \\ 
\vdots & \vdots & \vdots & \ddots & \vdots \\ 
0 & 0 & 0 & \cdots & 1 \\ 
-1 & 0 & 0 & \cdots & 0
\end{array}
\right) \,,  \label{pi}
\end{equation}
which satisfies the following properties $\Pi ^{N}=-I$ and $\Pi ^{T}=\Pi
^{-1}=-\Pi ^{N-1}\,$.

We have chosen the shift matrix $\Pi $ with a phase $e^{i\pi }=-1$ in order
to give odd parity to circuits winding around (see \cite{HG}).

From these definitions with $N=mn$, the matrix $A$ can be written as 
\begin{equation}  \label{A=A}
A=I\otimes U+\Pi \otimes X-\Pi ^{T}\otimes X^{T}+\Pi ^{m}\otimes Y-\left(
\Pi ^{T}\right) ^{m}\otimes Y^{T} \, .
\end{equation}

To block--diagonalize (\ref{A=A}) we introduce the $N\times N$ Fourier
matrix $F_{N}=F=\{f_{ij}\}$ defined as follows. Let $\omega =\omega
_{N}=e^{-\pi i/N}$, $i=\sqrt{-1}$, and note the following property

\begin{equation}
1+\omega ^{2}+\omega ^{4}+\cdots +\omega ^{2(N-1)}=0.  \label{omegas}
\end{equation}

We define 
\begin{equation}
f_{ij}:=\frac{1}{\sqrt{N}}\,\omega ^{(2i-1)\,j}\,.  \label{f}
\end{equation}

The Fourier matrix $F$ is a unitary matrix, $F\,F^{\dagger }=F^{\dagger
}\,F=I_{N}$, which diagonalizes the so--called class of $N\times N$
``circulant'' matrices. In particular, using (\ref{f}) and property (\ref
{omegas}), we have 
\begin{equation}
F^{-1}\,\Pi \,F={\rm diag}\,\{\omega ,\omega ^{3},\dots ,\omega
^{2N-1}\}\equiv D  \label{FPiF}
\end{equation}
and $F^{-1}\,\Pi ^{k}\,F=D^{k}$. In addition, taking the Hermitian conjugate
of (\ref{FPiF}), gives 
\[
\left( F^{-1}\,\Pi \,F\right) ^{\dagger }=F^{-1}\,\Pi ^{T}\,F={\bar{D}}\,, 
\]
where ${\bar{D}}={\rm diag}\,\{{\bar{\omega}},{\bar{\omega}}^{3},\dots ,{%
\bar{\omega}}^{2N-1}\}$.

Now, we apply the Fourier matrix ${\cal F}=F_{mn}\otimes I_{4}$ (${\cal F}%
^{-1}=F_{mn}^{-1}\otimes I_{4}$) in (\ref{A=A}) and use the multiplication
rule (\ref{mrule}) with (\ref{FPiF}) to obtain 
\begin{equation}
{\cal F}^{-1}\,A\,{\cal F}=I\otimes U+D\otimes X-{\bar{D}}\otimes
X^{T}+D^{m}\otimes Y-{\bar{D}}^{m}\otimes Y^{T}=\bigoplus_{j}R_{j}\,,
\label{FAF}
\end{equation}
where 
\begin{equation}
R_{j}=\left( 
\begin{array}{cccc}
0 & u_{6} & -u_{4}-{\bar{\omega}}^{j} & -u_{8} \\ 
-u_{6} & 0 & u_{7} & u_{3}+\omega ^{jm} \\ 
u_{4}+\omega ^{j} & -u_{7} & 0 & -u_{5} \\ 
u_{8} & -u_{3}-{\bar{\omega}}^{jm} & u_{5} & 0
\end{array}
\right) \,\,,  \label{R}
\end{equation}
and $j=2(r-1)n+2l-1$ with $r\in \{1,\dots ,m\}$ and $l\in \{1,\dots ,n\}$.

As a consequence, the determinant of $A$ can be written as 
\begin{equation}
\det A=\prod_{l=1}^{n}\prod_{r=1}^{m}\left| R_{j}\right| \,,  \label{detA}
\end{equation}
and evaluated by the Laplace method, 
\begin{equation}
\left| R_{j}\right| =\frac{1}{w_{1}^{2}}\left( a+2b\cos \left( \phi
_{r}+\psi _{l}^{0}\right) +2c\cos \varphi _{l}+2d\cos \left( \phi _{r}+\psi
_{l}^{+}\right) +2e\cos \left( \phi _{r}-\psi _{l}^{-}\right) \right) \,,
\label{detR}
\end{equation}
with 
\begin{equation}
\phi _{r}=\frac{(2r-1)\pi }{m}\,,\;\;\;\;\;\;\;\;\;\;\;\;\;\;\;\;\;\;\;\;\;%
\;\;\;\;\;\;\;\;\;\;\;\varphi _{l}=\frac{(2l-1)\pi }{n}\,,  \label{phiphi}
\end{equation}
$\psi _{l}^{0}=\varphi _{l}/m+\pi /m$, $\psi _{l}^{\pm }=(1\pm m)\varphi
_{l}/m+\pi /m$, and 
\begin{equation}
\begin{array}{lll}
a & = & w_{1}^{2}+w_{2}^{2}+w_{3}^{2}+w_{4}^{2}\,, \\ 
b & = & w_{4}-w_{2}\,w_{3}\,, \\ 
c & = & w_{3}-w_{2}\,w_{4}\,, \\ 
d & = & w_{3}\,w_{4}-w_{5}\,w_{6}\,, \\ 
e & = & w_{3}\,w_{4}-w_{7}\,w_{8}\,.
\end{array}
\label{abcde}
\end{equation}
Here, we have used $u_{\xi }=w_{\xi }/w_{1}$ and equation (\ref{ff}) to
obtain this form.


\section{The Free Energy}

\label{FE}

\setcounter{equation}{0} \setcounter{theorem}{0}

This section is concerned with the calculation of free energy $f=f(\beta )$,
given by (\ref{fe}), of a class of quantum spin systems satisfying (\ref{ff}%
). Two limits are required to be taken: the Trotter limit $m\rightarrow
\infty $ of the partition function $Z_{n,m}$ and the thermodynamic limit $%
n\rightarrow \infty $ of its logarithm. So far, we have obtained $Z_{n,m}$
exactly in Proposition \ref{Z=det} and equations (\ref{detA}) - (\ref{abcde}%
), for any value $n,m\in {\Bbb Z}_{+}$. In the limit, these equations
involve infinite products and the way we deal with them depends on whether $%
m $ or $n$ is taken to infinite. A third limit, $\lim_{\beta \rightarrow
\infty }f(\beta )$, will also be taken in order to get the ground state
energy.

We begin by studying the Trotter limit. Taylor expanding (\ref{w}) (with $%
J^{z}=0$) up to second order in $1/m$ and substituting the result into (\ref
{abcde}), gives 
\begin{equation}
\begin{array}{lllll}
&  & a & = & 2+\displaystyle\frac{2\beta ^{2}(h^{2}+\zeta ^{2}+\kappa ^{2})}{%
m^{2}}+{\cal O}\left( \displaystyle\frac{1}{m^{3}}\right) \,, \\ 
b & = & c & = & \displaystyle\frac{2\beta ^{2}h\kappa }{m^{2}}+{\cal O}%
\left( \displaystyle\frac{1}{m^{3}}\right) \,, \\ 
&  & d & = & -1\,, \\ 
&  & e & = & \displaystyle\frac{\beta ^{2}(\kappa ^{2}-\eta ^{2})}{m^{2}}+%
{\cal O}\left( \displaystyle\frac{1}{m^{3}}\right) \,,
\end{array}
\label{a-e}
\end{equation}
where $\eta =J^{x}-J^{y}$, $\kappa =J^{x}+J^{y}$ and $\zeta ^{2}=h^{2}+\eta
^{2}$. The value of $d$ is exact in view of (\ref{abcde}) and (\ref{w}).

We introduce a real valued function 
\begin{equation}
G(z):=\frac{1}{2}\left( a+c(z+{\bar{z}})\right) \,,  \label{Gz}
\end{equation}
and a complex valued function 
\begin{equation}
H(z):=z^{-(1-n+m)/m}\left( d+bz+ez^{2}\right)  \label{Hz}
\end{equation}
defined on the unit circle $|z|^{2}=1\,,$ and set 
\begin{eqnarray}
\Delta ^{2}(z) &=&G^{2}-H{\bar{H}}  \nonumber \\
&=&\frac{4\beta ^{2}}{m^{2}}\left( \frac{h^{2}+\zeta ^{2}+\kappa ^{2}}{2}+{%
h\kappa }(z+{\bar{z}})+\frac{\kappa ^{2}-\eta ^{2}}{4}(z^{2}+{\bar{z}}%
^{2})\right) +{\cal O}\left( \frac{1}{m^{3}}\right) \,,  \label{Deltaz}
\end{eqnarray}
in view of (\ref{a-e}). Note that $(G+\Delta )(G-\Delta )=H{\bar{H}}$.

Using these definitions, equation (\ref{detR}) can be rewritten as 
\begin{eqnarray}
w_{1}^{2}|R_{j}| &=&2\,G_{l}+H_{l}\,\omega _{r}+{\bar{H}}_{l}\,{\bar{\omega}}%
_{r}  \label{Rj} \\
&=&\frac{H_{l}}{\omega _{r}}\left( \frac{G_{l}+\Delta _{l}}{-H_{l}}-\omega
_{r}\right) \left( \frac{G_{l}-\Delta _{l}}{-H_{l}}-\omega _{r}\right) \,, 
\nonumber
\end{eqnarray}
where $\omega _{r}=e^{i\phi _{r}}$, $G_{l}=G\left( e^{i\varphi _{l}}\right) $
with $\phi _{r},\varphi _{l}$ as in (\ref{phiphi}) and similar expressions
holding for $H_{l}$ and $\Delta _{l}$.

Now, since $\{\omega _{r};r=1,\dots ,m\}$ are the roots of $-1$, we have 
\begin{equation}
\prod_{r=1}^{m}\left( x-\omega _{r}\right) =x^{m}+1\,,  \label{xwr}
\end{equation}
and the substitution of (\ref{Rj}) into (\ref{detA}) yields 
\begin{eqnarray}
w_{1}^{2nm}\,\det A &=&\prod_{l=1}^{n}\left( -H_{l}\right) ^{m}\left[ \left( 
\frac{G_{l}+\Delta _{l}}{-H_{l}}\right) ^{m}+1\right] \left[ \left( \frac{%
G_{l}-\Delta _{l}}{-H_{l}}\right) ^{m}+1\right]  \label{detAR} \\
&=&\prod_{l=1}^{n}\left\{ (-H_{l})^{m}+(-{\bar{H}}_{l})^{m}+(G_{l}+\Delta
_{l})^{m}+(G_{l}-\Delta _{l})^{m}\right\} \,,  \nonumber
\end{eqnarray}
where we have used $\prod_{r}(-\omega _{r})=1$.

To evaluate (\ref{detAR}), note that the asymptotic estimate 
\begin{equation}
\left( 1+\frac{c}{m}+{\cal O}\left( \frac{1}{m^{2}}\right) \right)
^{m}=e^{c}\left( 1+{\cal O}\left( \frac{1}{m}\right) \right)  \label{asympt}
\end{equation}
holds for all $c\in {\Bbb R}$. From (\ref{Hz}) and the fact that $m=kn+1$
(helical condition) for some positive integer $k$, we have 
\[
(-H_{l})^{m}=e^{-(k+1)n\varphi _{l}}\left( 1+{\cal O}\left( \frac{1}{m}%
\right) \right) \,, 
\]
which tends to $1$ or $-1$ depending on whether $k$ is odd or even. We
always take $k$ an odd number.

From equations (\ref{Gz}) and (\ref{Deltaz}) and asymptotic expansions (\ref
{a-e}), we have 
\begin{equation}
(G_{l}\pm \Delta _{l})^{m}=e^{\pm 2\delta _{l}}\left( 1+{\cal O}\left( \frac{%
1}{m}\right) \right) \,,  \label{GD}
\end{equation}
where $\delta _{l}=\delta (\varphi _{l})$ with 
\begin{equation}
\delta (\varphi )=\beta \left( h^{2}+(J^{x})^{2}+(J^{y})^{2}+2h\left(
J^{x}-J^{y}\right) \cos \varphi +2J^{x}J^{y}\cos 2\varphi \right) ^{1/2}\,.
\label{phil}
\end{equation}

This gives 
\begin{eqnarray}
w_{1}^{2nm}\,\det A &=&\prod_{l=1}^{n}\left( 2+e^{2\delta _{l}}+e^{-2\delta
_{l}}\right) \left( 1+{\cal O}\left( \frac{1}{m}\right) \right)  \nonumber \\
&=&\prod_{l=1}^{n}\left( 2\cosh {\delta _{l}}\right) ^{2}\left( 1+{\cal O}%
\left( \frac{1}{m}\right) \right)  \label{detAest} \\
&=&\exp \left( 2\sum_{l=1}^{n}\ln \left( 2\cosh \delta _{l}\right) \right)
\left( 1+{\cal O}\left( \frac{1}{m}\right) \right) \,,  \nonumber
\end{eqnarray}
and leads to the following expression for the partition function (see
Proposition \ref{Z=det}), 
\begin{equation}
Z_{n}=\lim_{m\rightarrow \infty }Z_{n,m}=\exp \left\{ \frac{n}{\pi }%
\int_{0}^{2\pi }d\varphi \,\,\ln \,\left[ 2\cosh \left( {\delta (\varphi )}%
\right) \right] \right\} \left( 1+o\left( \frac{1}{n}\right) \right) \,,
\label{Znn}
\end{equation}
where the limit is taken over the subsequence with $m=kn+1$, $k=1,3,5,\dots $%
. We have verified that the limit agrees with (\ref{Znn}) if taken over the
subsequence with $m=kn$, $k\in {\Bbb N}_{+}$, corresponding to toroidal
boundary condition. The remaining subsequences are expected to converge to
the same limit.

In approximating the Riemann sum by the Riemann integral in equation (\ref
{Znn}) we have used the fact that the integrand $g=\ln \left( 2\cosh \delta
\right) $ is a periodic and smooth function of $\varphi $ to get the sharper
than midpoint rule error estimate 
\[
\left| \frac{1}{n}\sum_{l}g(\varphi _{l})-\frac{1}{2\pi }\int_{0}^{2\pi
}g(\varphi )\,d\varphi \right| ={\cal O}\left( \frac{1}{n^{K}}\right) 
\]
for any $K\in {\Bbb N}_{+}$ (see Proposition $5.2$ of \cite{MFH}).

From equations (\ref{Zn1}) and (\ref{Znn}), the free energy function (\ref
{fe}) is thus given by 
\begin{equation}
f(\beta )=\frac{1}{\pi \beta }\int_{0}^{\pi }d\varphi \,\,\ln \,\left[
2\cosh \left( \delta (\varphi )\right) \right]  \label{fenergy}
\end{equation}
and the ground state energy, $e_{0}=\lim\limits_{\beta \rightarrow \infty
}f(\beta )$, by

\begin{equation}
e_{0}=\frac{1}{\pi }\int_{0}^{\pi }d\varphi \,\,\sqrt{%
h^{2}+(J^{x})^{2}+(J^{y})^{2}+2h(J^{x}-J^{y})\cos \varphi +2J^{x}J^{y}\cos
2\varphi }\,,  \label{grounde}
\end{equation}
where we have used the fact that $\delta (\varphi )$ is symmetric about $%
\varphi =\pi $.

This concludes the analysis of the period one $XY$ model in a transverse
magnetic field.


\section{$XY$ Model with Period $p=2$}

\label{XY-2}

\setcounter{equation}{0} \setcounter{theorem}{0}

The previous analysis of a homogeneous quantum spin system with period $p=1$
can be extended to arbitrary period. In this section we shall illustrate the
procedure fully for the period $p=2$ case.

For a periodic model with period $p$, all steps of Section \ref{TF} can be
repeated without modification. Since the purpose of that section was to
separate consecutive terms of the Hamiltonian (\ref{H}) we may adopt the
same odd--even decomposition (\ref{HH}), provided $p$ is an even number. We
also take $N=2pn$ and impose periodic boundary condition. The only
difference is that we now have $p$ matrices $\rho ^{(1)},\ldots ,\rho ^{(p)}$
, of the form (\ref{rhoM}) with distinct values $(\eta _{i},\kappa
_{i},\zeta _{i})$, $i=1,\ldots ,p$.

Section \ref{vertex} is also maintained with a minor modification. A vertex $%
v$ in the dual lattice $\Lambda _{n,m}^{\ast }$ will be now distinguished
according to its type labeled by $\{1,\ldots ,p\}$: the label increases
sequentially (modulo $p$) if one follows the northeast or southeast
directions as seem in Figure $7$.

\begin{figure}[ht]    
  \centering      
  \includegraphics[width=2.6066in,height=2.1793in]{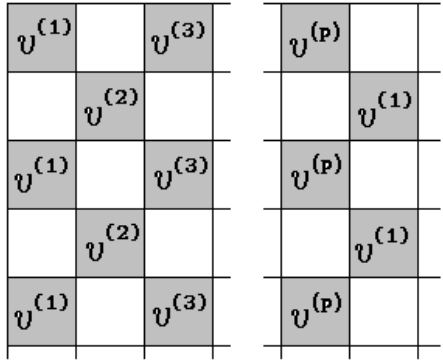}
  \caption{Dual lattice with  period $p$}
\end{figure}

We associate to each vertex $v^{(i)}$ of type $i$ a set $\{w_{1}^{(i)},%
\ldots ,w_{8}^{(i)}\}$ of weights given by (\ref{w}) with parameters $(\eta
_{i},\kappa _{i},\zeta _{i})$ and to each vertex (compatible) configuration $%
\xi =\{\xi _{v}\}$, a weight $w(\xi )=\prod_{v}w_{\xi _{v}}$ with $w_{\xi
_{v}}=w_{\xi _{v}}^{(i)}$ if $v$ is of the type $i$.

This completes the set up which allows Propositions \ \ref{Vertex} and \ref
{Z=det} to be holden for any period $p$ and we are now ready to evaluate the
determinant of the correspondent matrix $A$.

Up to this point there were no extra difficulties. The matrix $A$ is defined
by equation (\ref{m}) with the signal determined by the orientation of the
terminal graph ${\cal G}=(\Lambda _{n,m}^{\ast })^{T}$ as in Figure $6$. But
now the unit cells is composed by $p$ different types of vertices $\Box _{v}$
which leads to $4p\times 4p$ internal matrices. We shall write explicitly
these matrices, whose rows and columns are indexed by the label of Figure $8$%
, only for $p=2$ .

\begin{figure}[ht]    
  \centering      
  \includegraphics[width=1.5056in,height=1.4313in]{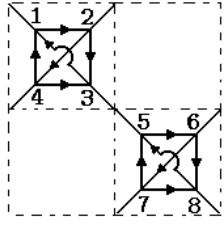}
  \caption{Unit cell for period $p=2$ }
\end{figure}

Similarly to what have been done in Section \ref{Det}, we set $u_{\xi
}\equiv u_{\xi }^{(1)}=w_{\xi }^{(1)}/w_{1}^{(1)}$ and $u_{\xi }^{\prime
}\equiv u_{\xi }^{(2)}=w_{\xi }^{(2)}/w_{1}^{(2)}$, and let 
\begin{equation}
V=\left( 
\begin{array}{cccccccc}
0 & u_{6} & -u_{4} & -u_{8} & 0 & 0 & 0 & 0 \\ 
-u_{6} & 0 & u_{7} & u_{3} & 0 & 0 & 0 & 0 \\ 
u_{4} & -u_{7} & 0 & -u_{5} & 1 & 0 & 0 & 0 \\ 
u_{8} & -u_{3} & u_{5} & 0 & 0 & 0 & 0 & 0 \\ 
0 & 0 & -1 & 0 & 0 & u_{6}^{\prime } & -u_{4}^{\prime } & -u_{8}^{\prime }
\\ 
0 & 0 & 0 & 0 & -u_{6}^{\prime } & 0 & u_{7}^{\prime } & u_{3}^{\prime } \\ 
0 & 0 & 0 & 0 & u_{4}^{\prime } & -u_{7}^{\prime } & 0 & -u_{5}^{\prime } \\ 
0 & 0 & 0 & 0 & u_{8}^{\prime } & -u_{3}^{\prime } & u_{5}^{\prime } & 0
\end{array}
\right) =\left( 
\begin{array}{cc}
U & X \\ 
-X^{T} & U^{\prime }
\end{array}
\right) \,,  \label{V}
\end{equation}
and 
\begin{equation}
Q=\left( 
\begin{array}{cc}
0 & 0 \\ 
X & 0
\end{array}
\right) \;,\qquad R=\left( 
\begin{array}{cc}
0 & Y \\ 
0 & 0
\end{array}
\right) \qquad {\rm and}\qquad S=\left( 
\begin{array}{cc}
0 & 0 \\ 
Y & 0
\end{array}
\right) \,,  \label{WZ}
\end{equation}
where $U,\,X$ and $Y$ is as in (\ref{U}) and (\ref{XY}). If $m=kn+1$, with $%
k $ an odd number, we recall that $\Lambda _{n,m}^{\ast }$ becomes helically
wound on a torus. The matrix $A$ can thus be written as (see Figure $9$ for
a self--explanation) 
\begin{equation}
A=I\otimes V+\Pi \otimes Q-\Pi ^{T}\otimes Q^{T}+\Pi ^{m}\otimes R-\left(
\Pi ^{T}\right) ^{m}\otimes R^{T}+\Pi ^{m+1}\otimes S-\left( \Pi ^{T}\right)
^{m+1}\otimes S^{T}\,,  \label{A2}
\end{equation}
where $I$ and $\Pi $ are the $nm\times nm$ identity and the forward shift.

\begin{figure}[ht]    
  \centering      
  \includegraphics[width=2.2935in,height=1.8299in]{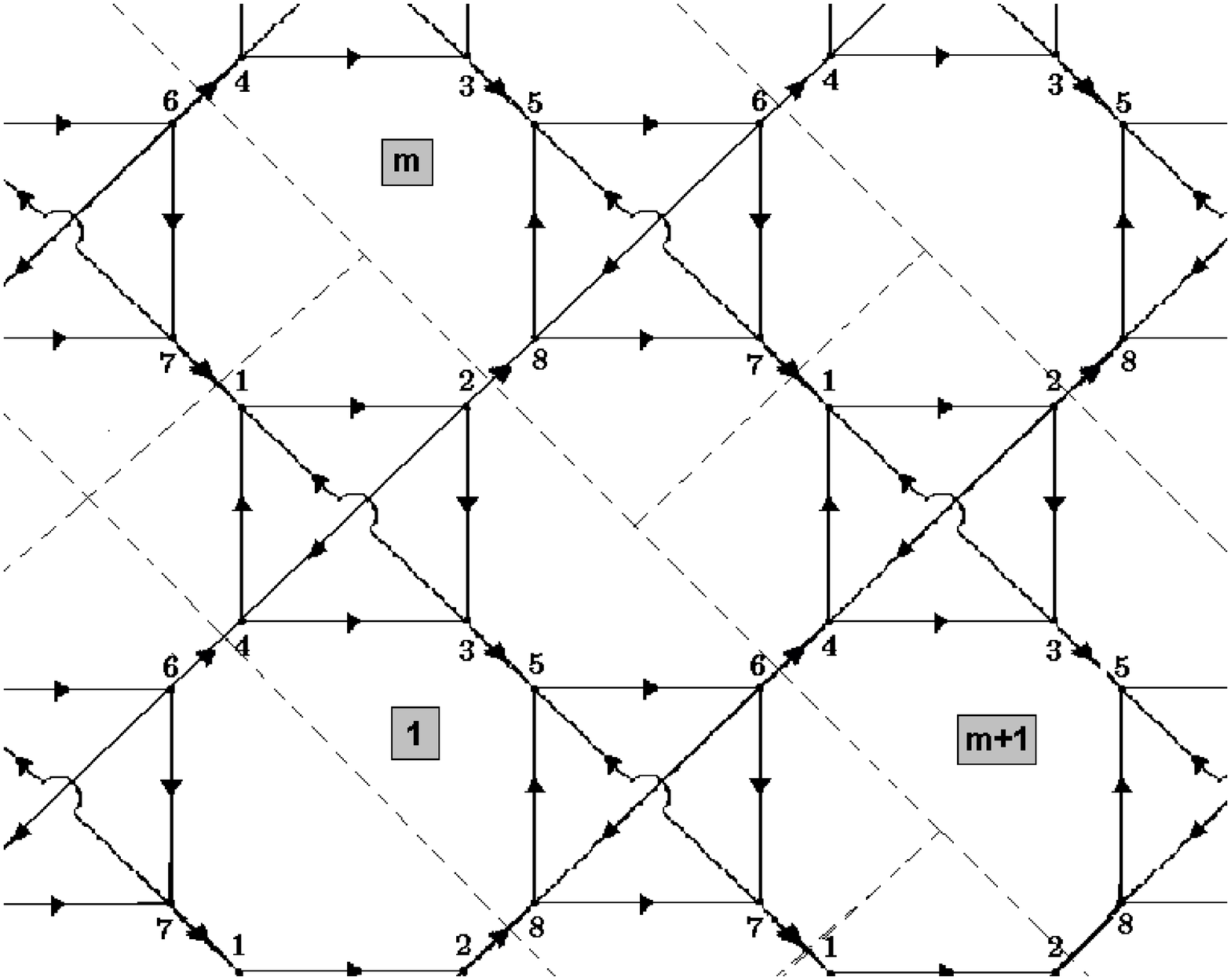}
  \caption{Numbering nearest   neighbors unit cells}
\end{figure}

Conjugating $A$ by the Fourier matrix ${\cal F}=F_{mn}\otimes I_{8}$ gives 
\begin{equation}
{\cal F}^{-1}\,A\,{\cal F}=\bigoplus_{j}S_{j}\equiv \bigoplus_{j}\left( 
\begin{array}{cc}
U & -M_{j}^{\dagger } \\ 
M_{j} & U^{\prime }
\end{array}
\right) \,,  \label{Sj}
\end{equation}
where $j=2(r-1)n+2l-1$, with $r\in \{1,\ldots ,m\}$, $l\in \{1,\ldots ,n\}$
and 
\begin{equation}
M_{j}=-X^{T}+\omega ^{j}X+\omega ^{j(m+1)}Y-{\bar{\omega}}^{jm}Y^{T}=\left( 
\begin{array}{cccc}
0 & 0 & -1 & 0 \\ 
0 & 0 & 0 & {\omega }^{j(m+1)} \\ 
\omega ^{j} & 0 & 0 & 0 \\ 
0 & -{\bar{\omega}}^{jm} & 0 & 0
\end{array}
\right) \text{,}  \label{Mj}
\end{equation}
with $\omega =e^{i\pi /(nm)}$. This implies 
\begin{equation}
\det A=\prod_{l=1}^{n}\prod_{r=1}^{m}\left| S_{j}\right| \,,  \label{detAA}
\end{equation}
where 
\begin{equation}
\left| S_{j}\right| =\left| U\right| \,\left| U^{\prime
}+M_{j}U^{-1}M_{j}^{\dagger }\right| \,.  \label{Sjj}
\end{equation}
Here, we used the fact that (\ref{Sj}) can be brought to block diagonal form
by a non--unitary transformation 
\[
\left( 
\begin{array}{cc}
1 & 0 \\ 
-M_{j}U^{-1} & 1
\end{array}
\right) \left( 
\begin{array}{cc}
U & -M_{j}^{\dagger } \\ 
M_{j} & U^{\prime }
\end{array}
\right) \left( 
\begin{array}{cc}
1 & U^{-1}M_{j}^{\dagger } \\ 
0 & 1
\end{array}
\right) =\left( 
\begin{array}{cc}
U & 0 \\ 
0 & U^{\prime }+M_{j}U^{-1}M_{j}^{\dagger }
\end{array}
\right) \,. 
\]

Using Laplace method and equation (\ref{u-zz}), we have $\left| U\right|
\,=u_{2}^{2}$ and

\[
U^{-1}=\frac{1}{u_{2}}\left( 
\begin{array}{cccc}
0 & -u_{5} & -u_{3} & u_{7} \\ 
u_{5} & 0 & -u_{8} & u_{4} \\ 
u_{3} & u_{8} & 0 & u_{6} \\ 
-u_{7} & -u_{4} & -u_{6} & 0
\end{array}
\right) \,, 
\]
which, in view of (\ref{Mj}) and (\ref{Sjj}), gives 
\begin{equation}
\left| S_{j}\right| =\frac{1}{u_{2}^{2}}\left| 
\begin{array}{cccc}
0 & u_{2}u_{6}^{\prime }-u_{6}{\omega }^{-j(m+1)} & -u_{2}u_{4}^{\prime
}-u_{3}{\omega }^{-j} & -u_{2}u_{8}^{\prime }+u_{8}{\omega }^{jm} \\ 
-u_{2}u_{6}^{\prime }+u_{6}{\omega }^{j(m+1)} & 0 & u_{2}u_{7}^{\prime
}-u_{7}{\omega }^{jm} & u_{2}u_{3}^{\prime }+u_{4}{\omega }^{j(2m+1)} \\ 
u_{2}u_{4}^{\prime }+u_{3}{\omega }^{j} & -u_{2}u_{7}^{\prime }+u_{7}{\omega 
}^{-jm} & 0 & -u_{2}u_{5}^{\prime }+u_{5}{\omega }^{j(m+1)} \\ 
u_{2}u_{8}^{\prime }-u_{8}{\omega }^{-jm} & -u_{2}u_{3}^{\prime }-u_{4}{%
\omega }^{-j(2m+1)} & u_{2}u_{5}^{\prime }-u_{5}{\omega }^{-j(m+1)} & 0
\end{array}
\right| \,.  \label{Sj-j}
\end{equation}

Using Laplace method and (\ref{u-zz}), equation (\ref{Sj-j}) can be written
as 
\begin{equation}
\begin{array}{lll}
(w_{1}w_{1}^{\prime })^{2}|S_{j}| & = & 2a+2b\cos (\xi _{j})+2c\cos (m\xi
_{j})+2d\cos ((m+1)\xi _{j}) \\ 
&  &  \\ 
&  & \qquad \qquad +2e\cos (2m\xi _{j})+2f\cos ((2m+1)\xi _{j})+2g\cos
(2(m+1)\xi _{j})\,,
\end{array}
\label{S-j}
\end{equation}
where $\xi _{j}=\pi j/nm=\phi _{r}+\varphi _{l}/m-\pi /m$. Defining 
\begin{equation}
\begin{array}{lll}
\Omega _{1} & = & w_{1}w_{1}^{\prime }+w_{2}w_{2}^{\prime }\,, \\ 
\Omega _{2} & = & w_{3}w_{3}^{\prime }+w_{4}w_{4}^{\prime }\,, \\ 
\Omega _{3} & = & w_{5}w_{6}^{\prime }+w_{6}w_{5}^{\prime }\,, \\ 
\Omega _{4} & = & w_{7}w_{8}^{\prime }+w_{8}w_{7}^{\prime }\,,
\end{array}
\label{Omegas}
\end{equation}
the $a$--$g$ constants are given by 
\begin{equation}
\begin{array}{lll}
2a & = & \Omega _{1}^{2}+\Omega _{2}^{2}+\Omega _{3}^{2}+\Omega
_{4}^{2}+2(w_{5}w_{6}w_{7}^{\prime }w_{8}^{\prime }+w_{7}w_{8}w_{5}^{\prime
}w_{6}^{\prime }-w_{1}w_{2}w_{1}^{\prime }w_{2}^{\prime
}-w_{3}w_{4}w_{3}^{\prime }w_{4}^{\prime })\,, \\ 
b & = & w_{5}w_{8}w_{6}^{\prime }w_{7}^{\prime }+w_{6}w_{7}w_{5}^{\prime
}w_{8}^{\prime }-w_{1}w_{4}w_{1}^{\prime }w_{4}^{\prime
}-w_{2}w_{3}w_{2}^{\prime }w_{3}^{\prime }\,, \\ 
c & = & \Omega _{2}\Omega _{3}-\Omega _{1}\Omega _{4}\,, \\ 
d & = & \Omega _{2}\Omega _{4}-\Omega _{1}\Omega _{3}\,, \\ 
e & = & (w_{3}w_{4}-w_{7}w_{8})(w_{3}^{\prime }w_{4}^{\prime }-w_{7}^{\prime
}w_{8}^{\prime })\,, \\ 
f & = & w_{6}w_{8}w_{5}^{\prime }w_{7}^{\prime }+w_{5}w_{7}w_{6}^{\prime
}w_{8}^{\prime }-w_{1}w_{3}w_{1}^{\prime }w_{3}^{\prime
}-w_{2}w_{4}w_{2}^{\prime }w_{4}^{\prime }\,, \\ 
g & = & (w_{3}w_{4}-w_{5}w_{6})(w_{3}^{\prime }w_{4}^{\prime }-w_{5}^{\prime
}w_{6}^{\prime })\,,
\end{array}
\label{abcdefg}
\end{equation}

Note that, in view of equations in (\ref{w}), $b=f$ and $g=1$ hold exactly.
This fact together with the trigonometric relations 
\[
\cos (\xi _{j})+\cos (2m\xi _{j}+\xi _{j})=2\cos ((m+1)\xi _{j})\,\cos (m\xi
_{j})\newline
\,, 
\]
and 
\[
\cos (2(m+1)\xi _{j})=2\cos ^{2}((m+1)\xi _{j})-1\,, 
\]
lead the right hand side of (\ref{Sj-j}) to be written as 
\begin{eqnarray}
(w_{1}w_{1}^{\prime })^{2}|S_{j}| &=&4\left( \cos ^{2}((m+1)\xi
_{j})-2B_{l}\cos ((m+1)\xi _{j})+C_{l}\right)  \label{Sj--j} \\
&=&4\left( \cos ((m+1)\xi _{j})-x_{l}^{+}\right) \left( \cos ((m+1)\xi
_{j})-x_{l}^{-}\right) \,,  \nonumber
\end{eqnarray}
where 
\[
\begin{array}{lll}
B_{l} & = & -\left( \displaystyle\frac{d}{4}+\displaystyle\frac{b}{2}\cos
(\varphi _{l})\right) \;, \\ 
&  &  \\ 
C_{l} & = & \displaystyle\frac{a-1}{2}+\displaystyle\frac{c}{2}\cos (\varphi
_{l})+\displaystyle\frac{e}{2}\cos (2\varphi _{l})
\end{array}
\]
(note that $\cos (m\xi _{j})=\cos (2(r+1)\pi +(2l-1)/n)=\cos (\varphi _{l})$%
) and 
\begin{eqnarray}
x_{l}^{\pm } &=&B_{l}\pm \sqrt{B_{l}^{2}-C_{l}}  \label{x+-} \\
&=&1+\frac{\beta ^{2}}{2m^{2}}\left\{ \kappa ^{2}+(\kappa ^{\prime
})^{2}+\eta ^{2}+(\eta ^{\prime })^{2}+(h+h^{\prime })^{2}+2(\kappa \kappa
^{\prime }-\eta \eta ^{\prime })\cos (\varphi _{l})\right\} \pm D_{l}+{\cal O%
}\left( \frac{1}{m^{4}}\right) \,,  \nonumber
\end{eqnarray}
where 
\begin{equation}
D_{l}^{2}=\frac{\beta ^{4}}{m^{4}}\left\{ \left[ \kappa \eta -\kappa
^{\prime }\eta ^{\prime }+(\kappa \eta ^{\prime }-\kappa ^{\prime }\eta
)\cos (\varphi _{l})\right] ^{2}+(h+h^{\prime })^{2}\left[ \kappa
^{2}+(\kappa ^{\prime })^{2}+2\kappa \kappa ^{\prime }\cos (\varphi _{l})%
\right] \right\} \,.  \label{D2}
\end{equation}

Here a word about the $1/m$ expansion is in order. Differently from the
period $p=1$ case, the constants (\ref{abcdefg}) of period $p=2$ have to be
computed up to order $1/m^{4}$ to pick up the relevant terms. This happens
because $B^{2}-C$, whose leading order term is written in equation (\ref{D2}%
), is ${\cal O}\left( {1}/{m^{4}}\right) $. So the square root $D_{l}$ has
the same order $1/m^{2}$ of $B_{l}-1$. Equation (\ref{Sj--j}) reduces the
present problem of evaluating $\det A$ to the one already dealt with in the
previous section.

Repeating the steps in (\ref{Gz}) -- (\ref{Rj}), we have 
\begin{equation}
(w_{1}w_{1}^{\prime })^{2}|S_{j}|=\frac{H_{l}^{2}}{\omega _{r}^{2}}\left( 
\frac{x_{l}^{+}+\Delta _{l}^{+}}{-H_{l}}-\omega _{r}\right) \left( \frac{%
x_{l}^{+}-\Delta _{l}^{+}}{-H_{l}}-\omega _{r}\right) \left( \frac{%
x_{l}^{-}+\Delta _{l}^{-}}{-H_{l}}-\omega _{r}\right) \left( \frac{%
x_{l}^{-}-\Delta _{l}^{-}}{-H_{l}}-\omega _{r}\right) \,,  \label{SSjj}
\end{equation}
where $H_{l}=(\omega _{l})^{-(1-n+m)/m}$ and $\left( \Delta _{l}^{\pm
}\right) ^{2}=(x_{l}^{\pm })^{2}-H_{l}{\bar{H}}_{l}=(x_{l}^{\pm })^{2}-1$
which, in view of (\ref{D2}), can be estimated by 
\begin{equation}
\left( \Delta _{l}^{\pm }\right) ^{2}=\frac{\beta ^{2}}{m^{2}}\left\{ \kappa
^{2}+(\kappa ^{\prime })^{2}+\eta ^{2}+(\eta ^{\prime })^{2}+(h+h^{\prime
})^{2}+(\kappa \kappa ^{\prime }-\eta \eta ^{\prime })\cos (\varphi
_{l})\right\} \pm 2D_{l}+{\cal O}\left( \frac{1}{m^{4}}\right) \,.
\label{Delta+-}
\end{equation}
Note that $\Delta _{l}^{\pm }={\cal O}\left( {1}/{m}\right) $.

Now, repeating the steps from (\ref{xwr}) to (\ref{detAest}), leads (\ref
{detAA}) to be written as 
\begin{equation}
(w_{1}w_{1}^{\prime })^{2}\det A=\exp \left( 2\sum_{l=1}^{n}\left[ \ln
\left( 2\cosh {\delta _{l}^{+}}\right) +\ln \left( 2\cosh {\delta _{l}^{-}}%
\right) \right] \right) \left( 1+{\cal O}\left( \frac{1}{m}\right) \right)
\,,  \label{detAAA}
\end{equation}
where $\delta _{l}^{\pm }\equiv \delta ^{\pm }(\varphi _{l})=\displaystyle%
\lim_{m\rightarrow \infty }m\,\Delta _{l}^{\pm }/2$ which, when expressed in
terms of the couplings and magnetic fields, is given by 
\[
\left( \delta ^{\pm }(\varphi )\right) ^{2}=\frac{\beta ^{2}}{2}\left(
(J_{1}^{x})^{2}+(J_{2}^{x})^{2}+(J_{1}^{y})^{2}+(J_{2}^{y})^{2}+2(J_{1}^{x}J_{2}^{y}+J_{1}^{y}J_{2}^{x})\cos \varphi \right) \qquad \qquad \qquad \qquad \qquad 
\]
\[
\qquad \qquad \qquad \pm \frac{\beta ^{2}}{2}\left\{ \left[
(J_{1}^{x})^{2}-(J_{2}^{x})^{2}-(J_{1}^{y})^{2}+(J_{2}^{y})^{2}+2(J_{1}^{x}J_{2}^{y}-J_{1}^{y}J_{2}^{x})\cos \varphi %
\right] ^{2}\right. 
\]
\[
\qquad \qquad \qquad \left. +(h_{1}+h_{2})^{2}\left[
(J_{1}^{x}+J_{1}^{y})^{2}+(J_{2}^{x}+J_{2}^{y})^{2}+2(J_{1}^{x}J_{2}^{x}+J_{1}^{x}J_{2}^{y}+J_{1}^{y}J_{2}^{x}+J_{1}^{y}J_{2}^{y})\cos \varphi %
\right] \right\} ^{1/2}\,. 
\]

The free energy function (\ref{fe}) for $p=2$ periodic systems is thus given
by 
\begin{equation}
f_{2}(\beta )=f^{+}(\beta )+f^{-}(\beta )\,,  \label{freeenerg}
\end{equation}
with 
\begin{equation}
f^{\pm }(\beta )=\frac{1}{2\pi \beta }\int_{0}^{\pi }d\varphi \,\ln \,\left[
2\cosh \left( {\delta ^{\pm }(\varphi )}\right) \right] \,.  \label{free}
\end{equation}
Note that the total size $N$ of the system is now equal to $4n$ and this
explain why (\ref{free}) is formally $1/2$ of (\ref{fenergy}).

It is interesting to observe that, if we set $J_{1}^{x}=J_{2}^{x}=J^{x}$, $%
J_{1}^{y}=J_{2}^{y}=J^{y}$ and $h_{1}=h_{2}=h$, we have 
\[
\left( \delta ^{\pm }(\varphi )\right) ^{2}=\beta \left(
(J^{x})^{2}+(J^{y})^{2}+2J^{x}J^{y}\cos \varphi \pm 2h(J^{x}+J^{y})\left|
\cos ({\varphi }/{2})\right| \right) \,, 
\]
and this gives (since $\delta ^{\pm }(\varphi )$ is symmetric about $\varphi
=\pi $) 
\begin{equation}
f_{2}(\beta )=\frac{1}{\pi \beta }\left( \int_{0}^{\pi /2}d\tau \,\ln \,%
\left[ 2\cosh \left( {\delta ^{+}(2\tau )}\right) \right] +\int_{0}^{\pi
/2}d\tau \,\ln \,\left[ 2\cosh \left( {\delta ^{-}(2\tau )}\right) \right]
\right) \,,  \label{freeenrgy}
\end{equation}
whose combination of the two terms yields (\ref{fenergy}).


\end{document}